\documentclass[onecolumn,superscriptaddress]{revtex4}
\usepackage{graphicx,epsfig}% Include figure files

\DeclareGraphicsExtensions{. jpg,. pdf, . mps, .png, .eps, . ps, . EPS}
\usepackage{amsmath}
\usepackage{color}
\begin{document}
\newcommand{\beq}{\begin{equation}}
\newcommand{\eeq}{\end{equation}}
\newcommand{\bea}{\begin{eqnarray}}
\newcommand{\eea}{\end{eqnarray}}
\newcommand{\gt}{\tilde{g}}
\newcommand{\mt}{\tilde{\mu}}
\newcommand{\et}{\tilde{\varepsilon}}
\newcommand{\ct}{\tilde{C}}
\newcommand{\bt}{\tilde{\beta}}

\newcommand{\avg}[1]{\langle{#1}\rangle}
\newcommand{\Avg}[1]{\left\langle{#1}\right\rangle}
\newcommand{\cor}[1]{\textcolor{red}{#1}}

\title{Control of Multilayer Networks}
\author{Giulia Menichetti}
\affiliation{Department of Physics and Astronomy and INFN Sez. Bologna, Bologna University, Viale B. Pichat 6/2 40127 Bologna, Italy}
\email{giulia.menichetti2@unibo.it}
\author{Luca Dall'Asta}
\affiliation{Department of Applied Science and Technology, DISAT, Politecnico di Torino, Corso Duca degli Abruzzi 24, 10129 Torino, Italy}
\affiliation{Collegio Carlo Alberto, Via Real Collegio 30, 10024 Moncalieri, Italy}
\thanks{corresponding author}
\email{luca.dallasta@polito.it}
\author{Ginestra Bianconi}
\affiliation{School of Mathematical Sciences, Queen Mary University of London, London E1 4NS, United Kingdom}
\email{ginestra.bianconi@gmail.com}

\begin{abstract}

{ \bf The controllability of a network  is a theoretical problem  of relevance in a variety of contexts ranging from   financial markets to the brain.
Until now, network controllability has been characterized only on isolated networks, while the vast majority of complex systems  are formed by multilayer networks. Here we build a theoretical framework for the linear  controllability of multilayer networks by mapping the problem into a combinatorial matching problem. We found that  correlating the external signals in the different layers can significantly reduce the multiplex network robustness to node removal, as it can be seen in conjunction with a hybrid phase transition occurring in interacting Poisson networks. 
Moreover we observe that multilayer networks can stabilize the fully controllable multiplex network configuration that can be stable also when the full controllability of the single network is not stable.}

\end{abstract}
\maketitle

%\tableofcontents

%\section*{Introduction}
Most of the real networks are not isolated but interact with each other forming multilayer structures \cite{PhysReports, review2}. For example, banks are linked to each other by different types of contracts and relationships, gene regulation in the cell is mediated by the different types of interactions between different kinds of molecules, brain data are described by multilayer brain networks. Studying the controllability properties of these networks is important for assessing the risk of a financial crash \cite{Caldarelli,Caldarelli2}, for drug discovery \cite{drug_discovery} and for characterizing brain dynamics \cite{Bullmore,Bonifazi,brain,Makse,Remondini}. Therefore the controllability of multilayer networks is a problem of fundamental importance for a large variety of applications.

Recently, linear \cite{Lin74,Liu,correlations,Ruths,Reka,PRL,con_vicsek,Spectrum,Control_centrality} and non-linear \cite{slotine,con_pinning,con_sorrentino,con_boccaletti,DiBernardo1,DiBernardo2,con_grebogi,con_lai,Motter,Motter1,Lai2} approaches are providing new scenarios for the characterization of the controllability of single complex  networks.  In particular, in the seminal paper by Liu et al. \cite{Liu} the structural controllability of complex  networks has been addressed by mapping this problem into a Maximum Matching Problem that can be studied using statistical mechanics techniques \cite{Lenka,Altarelli,Cavity,Zecchina,Weigt,Mezard}. Other works  approach the related problem of network observability \cite{observability}, or target control \cite{target} which focuses on controlling just a subset of the nodes.  Despite the significant interest in network controllability, all linear and non-linear approaches for the controllability of networks are still restricted to single networks while it has been recently found that the multiplexity of networks can have profound effects on the dynamical processes taking place on them \cite{Havlin1, Havlin3,Doro1,BD,Goh,Diffusion}. For example, percolation processes that usually present continuous phase transitions on single networks can become discontinuous on such structures \cite{Havlin1, Havlin3,Doro1,BD,Goh} and are characterized by large avalanches of disruption events.  

Here, we consider the elegant framework of structural controllability \cite{Liu} and investigate how the multilayer structure of networks can affect their controllability. 
We focus on multiplex networks, which are multilayer networks in which the same set of nodes are connected by different types of interactions. Multiplex network controllability is studied under the assumption that input nodes are the same in all network layers, thus mimicking the situation in which input nodes can send different signals in the different layers of the multiplex but the position of the external signals in the layers is correlated. 

We show that controlling the dynamics of multiplex networks is more costly than controlling single layers taken in isolation. Moreover, the controllability of multiplex networks displays unexpected new phenomena. In fact these networks can become extremely sensible to damage in conjunction with a discontinuous phase transition characterized by a jump in the number of input points (driver nodes). A careful investigation of this phase transition reveals that this is a hybrid phase transition with a square root singularity, therefore in the same universality class of the emergence of the mutually connected component in multiplex networks \cite{Havlin1,Doro1,PhysReports}. 
The number of driver nodes in the multiplex network is in general higher than the number of driver nodes in the single layers taken in isolation. Nevertheless the degree correlations between low-degree nodes in the different layers can affect the controllability of the multiplex network and modulate the number of its driver nodes. Moreover, a fully controllable configuration can be stable in a multilayer network even if it is not stable in the isolated networks that form the multilayer structure.

%%%%%%%%%%%%%%%%%%%%%%%%%%%%%%%%%%%%%%%%%%%%%%%%%%%%%%%%%%%
\begin{figure}
\begin{center}
{\includegraphics[width=11cm]{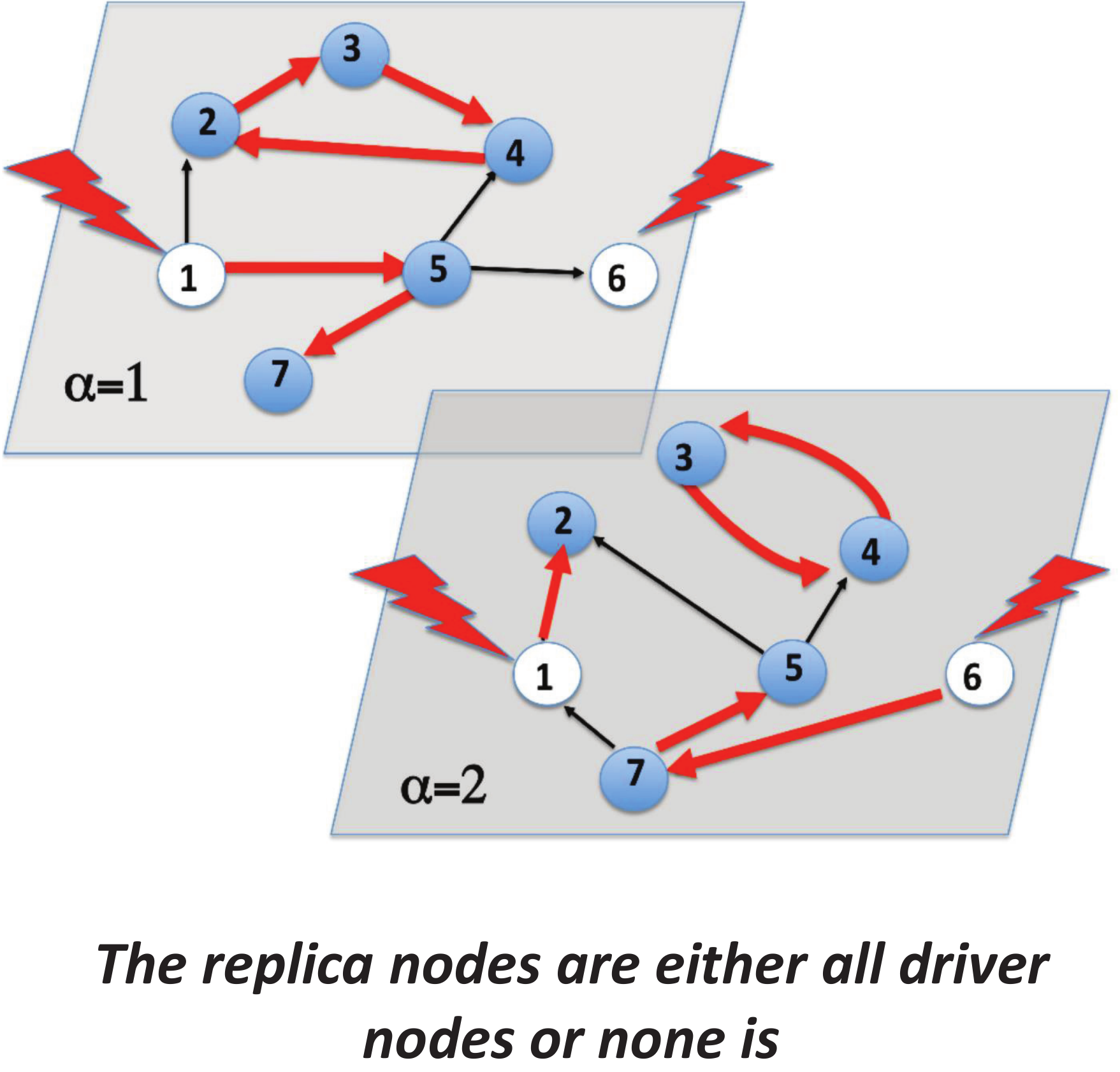}}
\end{center}
\caption{{\bf Control of a multiplex network}.  The controllability of a duplex network (multiplex with $M=2$ layers) can be mapped to a Maximum Matching Problem in which the unmatched nodes (indicated with a white circle) are the driver nodes of the duplex network. Here we have indicated with red thick links the matched links and by black thin links the unmatched links.} 
\label{fig1}
\end{figure}

\section*{RESULTS}

We consider  multiplex networks \cite{PhysReports} in which  every node $i=1,2,\ldots, N$ has a replica node $(i,\alpha)$ in each layer $\alpha$ and every layer is formed by a directed network between the corresponding replica nodes. We assume that each replica node $(i,\alpha)$ is characterized by a different dynamical variable $x_i^\alpha\in \mathrm{R}$ and that each layer is characterized by a possibly different dynamical process. We  consider for simplicity a duplex, i.e a multiplex formed by two layers $\{A,B\}$ where each layer $\alpha \in \{A,B\}$ is a directed network. The state of the network at time $t$ is governed by a linear dynamical system 
\bea
\frac{d{\bf X}(t)}{dt}={\cal G}{\bf X}+{\cal K}{\bf u},
\label{dyn}
\eea
in which the $2N$-dimensional vector ${\bf X}(t)$ describes the dynamical state of each replica node, i.e. $X_{i}=x_i^{A}$ and $X_{N+i}=x_i^{B}$ for $i=1,2,\ldots, N$.
The matrix ${\cal G}$ is a $2N\times 2N$ (asymmetric) matrix and ${\cal K}$ is a $2N\times P$ matrix. They have the following block structure
\bea
{\cal G}=\left(\begin{array}{cccc} g^{A}&0 \\ 0& g^{B}  \end{array}\right), \ \ \ {\cal K}=\left(\begin{array}{cc}K^{A}& 0 \\ 0& K^{B}\end{array}\right),
\label{bB}
\eea 
in which $g^{\alpha}$ are the $N\times N$ matrices describing the directed weighted interactions within the layers and  ${ K}^{\alpha}$ are the $N\times P^{\alpha}$ matrices describing the coupling between the nodes of each layer $\alpha$ and $P^{\alpha}\leq N$ external signals. The latter are represented by a vector ${\bf u}(t)$ of elements $u_{\gamma}$ and $\gamma=1,2\ldots P=P^{A}+P^{B}$. 
Here we consider the concept of structural controllability \cite{Lin74,Liu} that guarantees the controllability of a networks for any choice of the non-zeros entries of ${\cal G}$ and ${\cal K}$, except for a variety of zero Lebesgue measure in the parameter space. Therefore each layer of the duplex networks can be structurally controlled by identifying a minimum number of {\em driver nodes}, that are controlled nodes which do not share input vertices. 
If different replicas of the same node can be independently controlled, then the controllability properties of the multiplex network factorize and each layer can be studied as if was taken in isolation \cite{slotine,Lin74,Liu,PRL}. 
Liu et al. \cite{Liu} showed that in a single layer the minimum set of driver nodes can be found by mapping the problem into a matching problem. 
%In fact the Minimum Input Theorem states that the minimum set of driver nodes that guarantees the full structural controllability of a single network is the set of unmatched nodes in a maximum matching of the same directed network.
In real multiplex networks however nodes are usually univocally defined and share common properties across different layers, therefore we make the assumption that {\em each node of the duplex network is either a driver node in each layer or it is not a driver node in any layer}.   
The problem of finding the driver nodes of the duplex network can be thus mapped into a maximum matching problem in which every node has at most one matched incoming link and at most one matched outgoing link, with the constraint that two replica nodes either have no matched incoming links on each layer or have one matched incoming link in each layer (see Figure $\ref{fig1}$).
This problem can be studied,  using statistical mechanics techniques, such as the cavity method and the Belief Propagation (BP) algorithm.  Following \cite{Liu,PRL}, we consider the variables $s_{ij}^{\alpha}=1,0$ indicating respectively if the directed link from node $(i,\alpha)$ to node $(j,\alpha)$ in layer $\alpha=A,B$ is matched or not. In order to have a matching in each layer of the duplex the following constraints have always to be satisfied
\bea
\sum_{j\in \partial_{+}^{\alpha}i}s_{ij}^{\alpha}\leq 1,\ \ \ \sum_{i\in \partial_{-}^{\alpha}j}s_{ij}^{\alpha} \leq 1.
\label{uno}
\eea 
where $\partial_{+}^{\alpha}i$ is the set of replica nodes $(j,\alpha)$ in layer $\alpha$ that are reached by directed links from $(i,\alpha)$ and $\partial_{-}^{\alpha}j$ is the set of replica nodes $(i,\alpha)$ in layer $\alpha$ that point to $(j,\alpha)$. 
In addition, we impose that the driver nodes in the two layers (the unmatched nodes) are replica nodes, i.e. 
\bea
\sum_{i\in \partial_{-}^{A}j}s_{ij}^{A}=\sum_{i\in \partial_{-}^{B}j}s_{ij}^{B}.
\label{due}
\eea
In this formalism, computing the maximum matching corresponds to minimize an energy function  $E=N_D=N n_D$ where $N_D$ is the number of unmatched replica nodes associated to each matching. The energy $E$ for a given matching, can be expressed in terms of the variables $s_{ij}$ as   
\bea\label{ene}
E&=&\sum_{\alpha}\sum_j\left(1-\sum_{i\in \partial_{-}^{\alpha}j}s_{ij}^{\alpha}\right)\label{E}.
\eea
In order to study this novel statistical mechanics problem, we derived the BP equations \cite{Mezard,Cavity,Weigt} (see Methods and Supplementary Material)   valid in the locally tree-like approximation, as described for the case of a single network problem in \cite{Lenka,Altarelli,Zecchina,Liu,PRL}.

\begin{figure}
\center
\includegraphics[width=16cm]{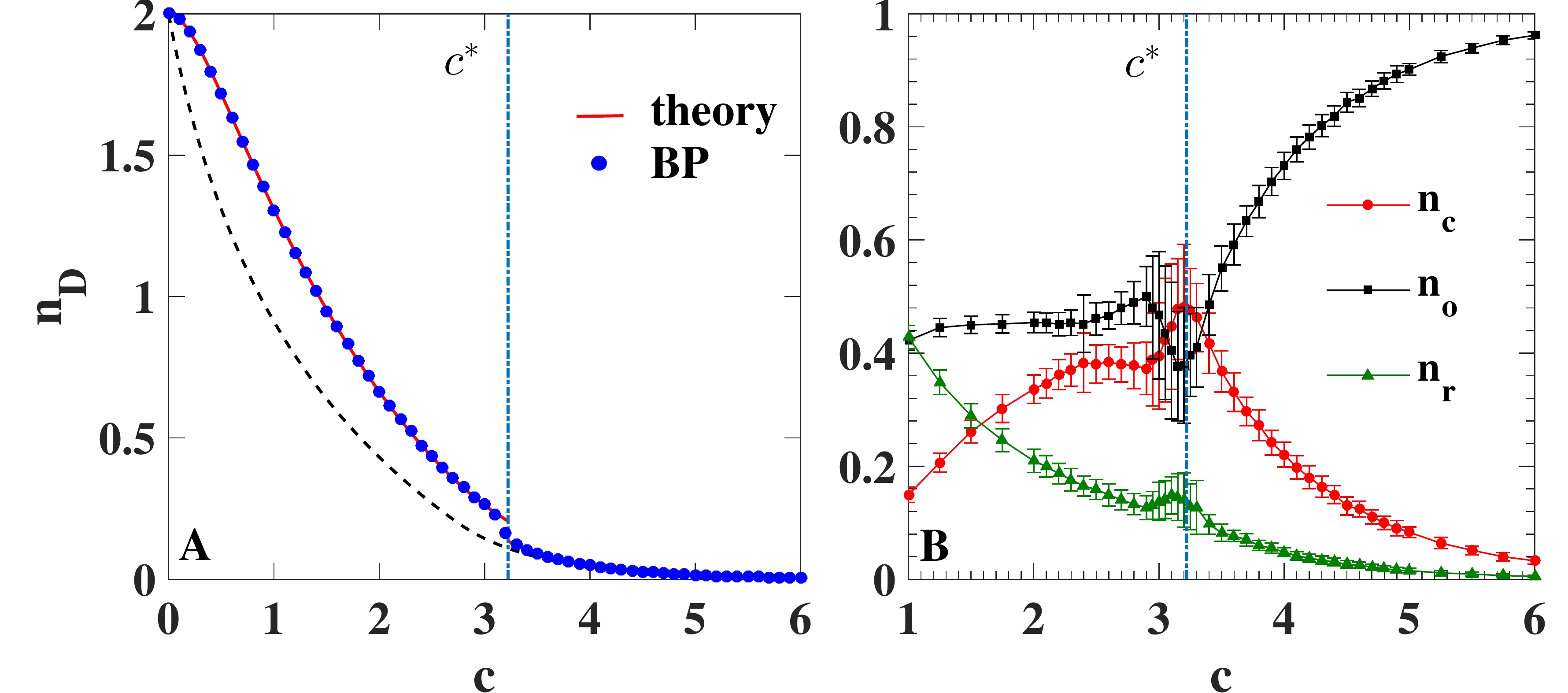}
\caption{ {\bf Controllability of  Poisson duplex networks with average degrees $\Avg{k^{A,in}}=\Avg{k^{A,out}}=\Avg{k^{B,in}}=\Avg{k^{B,out}}=c$.} In panel A the fraction $n_D$ of  driver nodes  in a Poisson duplex network  with $\Avg{k^{A,in}}=\Avg{k^{A,out}}=\Avg{k^{B,in}}=\Avg{k^{B,out}}=c$, plotted as a function of the average degree $c$.   The points indicate the average BP results  obtained over 5  single realizations of the Poisson duplex  networks with average degree $c$ and  $N=10^4$, the solid line is the theoretical expectation (the error bar, indicating the interval of one standard deviation from the mean, is always smaller or comparable to marker size). The dashed line represents twice the density of driver nodes for a single  Poisson network with the same average degree.\\
In panel B the densities  $n_c, n_r$ and $n_o$ respectively of critical redundant and ordinary nodes are shown  as functions of $c$ for the same type of duplex networks with $N=10^3$, where each point is the average over 100 different instances.\\
In both panels the dot-dashed  vertical line indicates the phase transition average degree $c^*=3.222326106\ldots$.}
\label{fig2}
\end{figure}

%%%%%%%%%%%%%%%%%%%%%%%%%%%%%%%%%%%%%%%%%%%%%%%%%%%%%%%%%%%
\begin{figure}
\center
{\includegraphics[width=8cm, angle=-90]{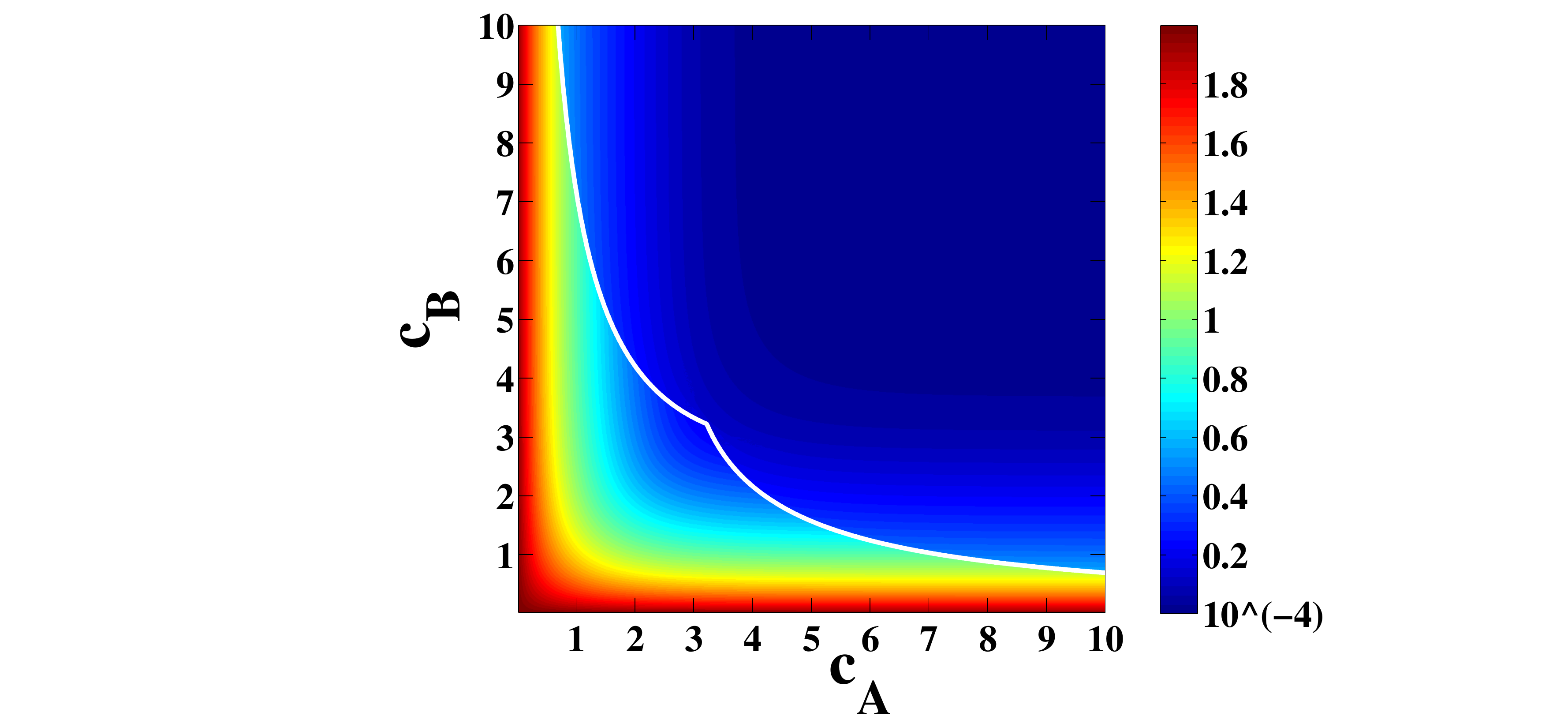}}
\caption{{\bf Phase diagram of the controllability  for a Poisson duplex networks with average  degrees $\Avg{k^{A,in}}=\Avg{k^{A,out}}=c_A$ and $\Avg{k^{B,in}}=\Avg{k^{B,out}}=c_B$.} The color code indicates the density  of driver nodes  $n_D=E/N$ in the multiplex network. } 
\label{fig3}
\end{figure}
\section*{DISCUSSION}

The controllability of multiplex networks displays a rich phenomenology, coming from the interplay between the dynamical and the structural properties of multiplex networks. Here we characterize  the controllability of multiplex networks with different degree distribution and with tunable level of structural correlations.

\paragraph*{Phase transition in Poisson duplex networks--}
We consider duplex networks in which the two layers are realizations of uncorrelated directed random graphs characterized by Poisson distributions for in-degree and out-degree with same average value $c$, i.e. $\avg{k^{A}_{in}}=\avg{k^{A}_{out}}=\avg{k^{B}_{in}}=\avg{k^{B}_{out}}=c$.  In Figure~\ref{fig2}A we report the average rescaled number of driver nodes $n_D$ as function of the average degree $c$ computed from the solutions of Eqs.(\ref{bp}) on single instances   and from the graph ensemble analysis. A comparison with two independent layers with the same topological properties  shows that the controllability of a duplex network is in general more demanding in terms of number of driver nodes than the controllability of independent single layers, in particular for low average degrees. In addition, a discontinuity in the number of driver nodes at $c=c^{\star}=3.2223\ldots$ marks a change in the controllability properties of duplex networks that is not observed in uncoupled networks. This is due to a structural change in the solution of the matching problem, in which a finite density of zero valued cavity fields emerges. A careful investigation (see Supplementary Material ) reveals that this is a hybrid phase transition with a square root singularity, therefore in the same universality class of the emergence of the mutually connected component in multiplex networks \cite{Havlin1,Doro1,PhysReports}. 

In correspondence to this phase transition the network responds non trivially to perturbations. This is observed by performing a numerical calculation of the robustness of the networks.
Following \cite{Liu} we classify the nodes into three categories: {\em critical nodes}, {\em redundant nodes } and {\em ordinary nodes}. When a critical node is removed from the (multiplex) network, controllability is sustained at the cost of increasing the number of driver nodes. If the number of driver nodes decreases or is unchanged, the removed nodes are classified as redundant and ordinary respectively. 
Figure \ref{fig2}B shows that the fraction of critical nodes reaches a maximum at the transition, revealing an increased fragility of the duplex network to random damage with respect to single layers. While an abrupt change in the number of driver nodes can result from a small change in the network topology, it is important to stress that the non-monotonic behavior of these quantities around the critical average degree value could be interpreted as a precursor of the discontinuity. 

In a duplex network formed by directed Poisson random graphs with different average degree in the two layers (i.e. $\avg{k^{\alpha}_{in}}=\avg{k^{\alpha}_{out}}=c_\alpha$) a similar discontinuous phase transition is observed (see Figure \ref{fig3}). Nevertheless we checked that this discontinuous phase transition is not occurring for every multiplex  network structure (see Supplementary Material).
%%%%%%%%%%%%%%%%%%%%%%%%%%%%%%%%%%%%%%%%%%%%%%%%%%%%%%%%%%%
\begin{figure}
\begin{center}
{\includegraphics[width=\columnwidth]{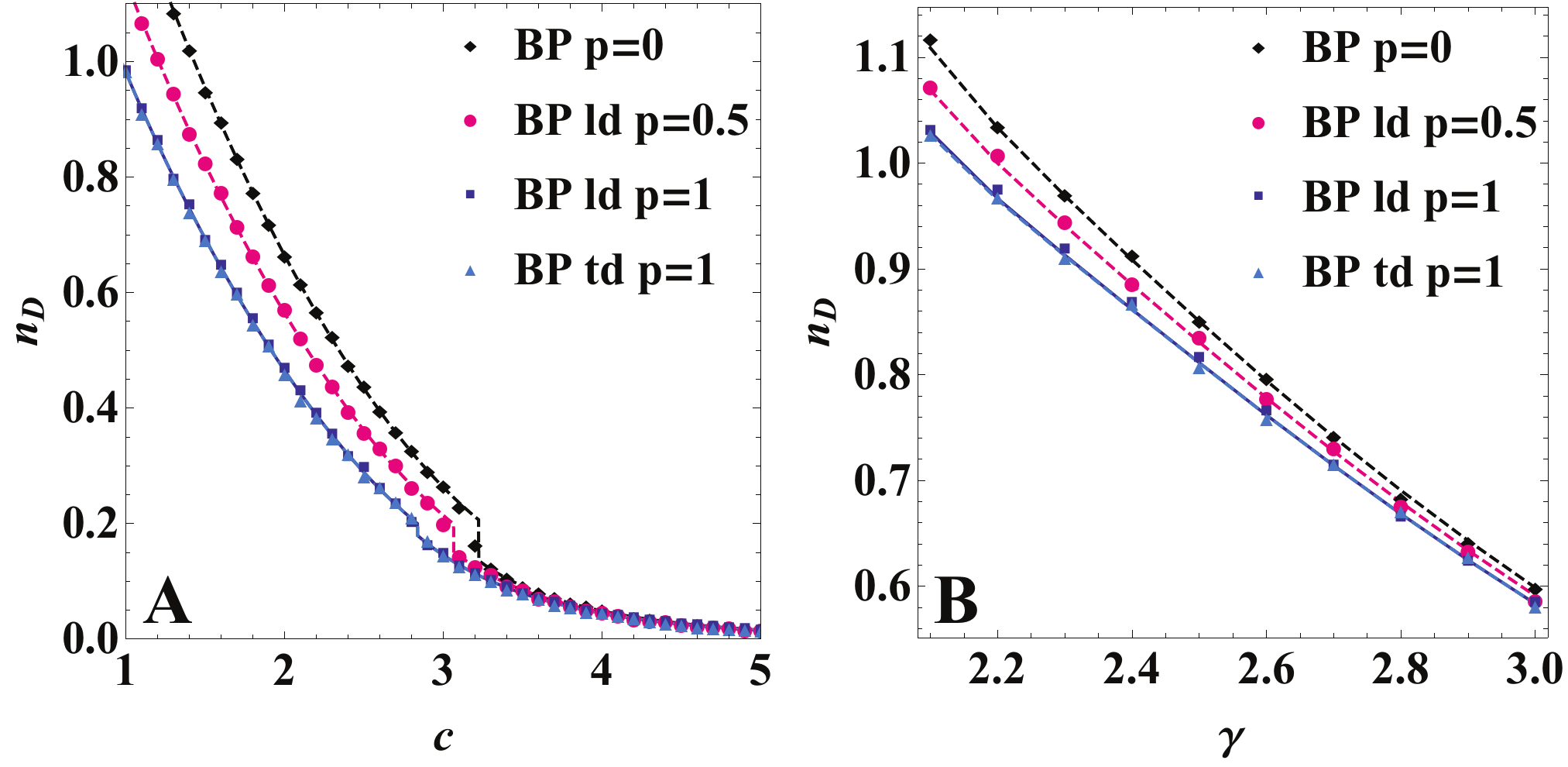}}
\end{center}
\caption{{\bf The effect of the degree correlation between replica nodes in different layers on the controllability of multiplex networks.} Correlations between the low in-degrees (ld) and correlations between any in-degree node (td),  parametrized by  $p$, affect the fraction of driver nodes in the network $n_D$, both in the case of Poisson networks with the same in and out average degree $c$ across the two layers (Panel A) and in the case of scale-free networks with the same in and out degree distribution across the layers, given by $P(k)\propto k^{-\gamma}$ and minimum in/out degree $1$  (Panel B). When $p=0$ there is no degree correlation between replica nodes in different layers. The BP data are shown for networks with $N=10^4$, and are  averaged 5 times for panel A and 20 times for panel B.} 
\label{fig:corr}
\end{figure}
%%%%%%%%%%%%%%%%%%%%%%%%%%%%%%%%%%%%%%%%%%%%%%%%%%%%%%%%%%
\begin{figure}
\begin{center}
{\includegraphics[width=8cm, angle=-90]{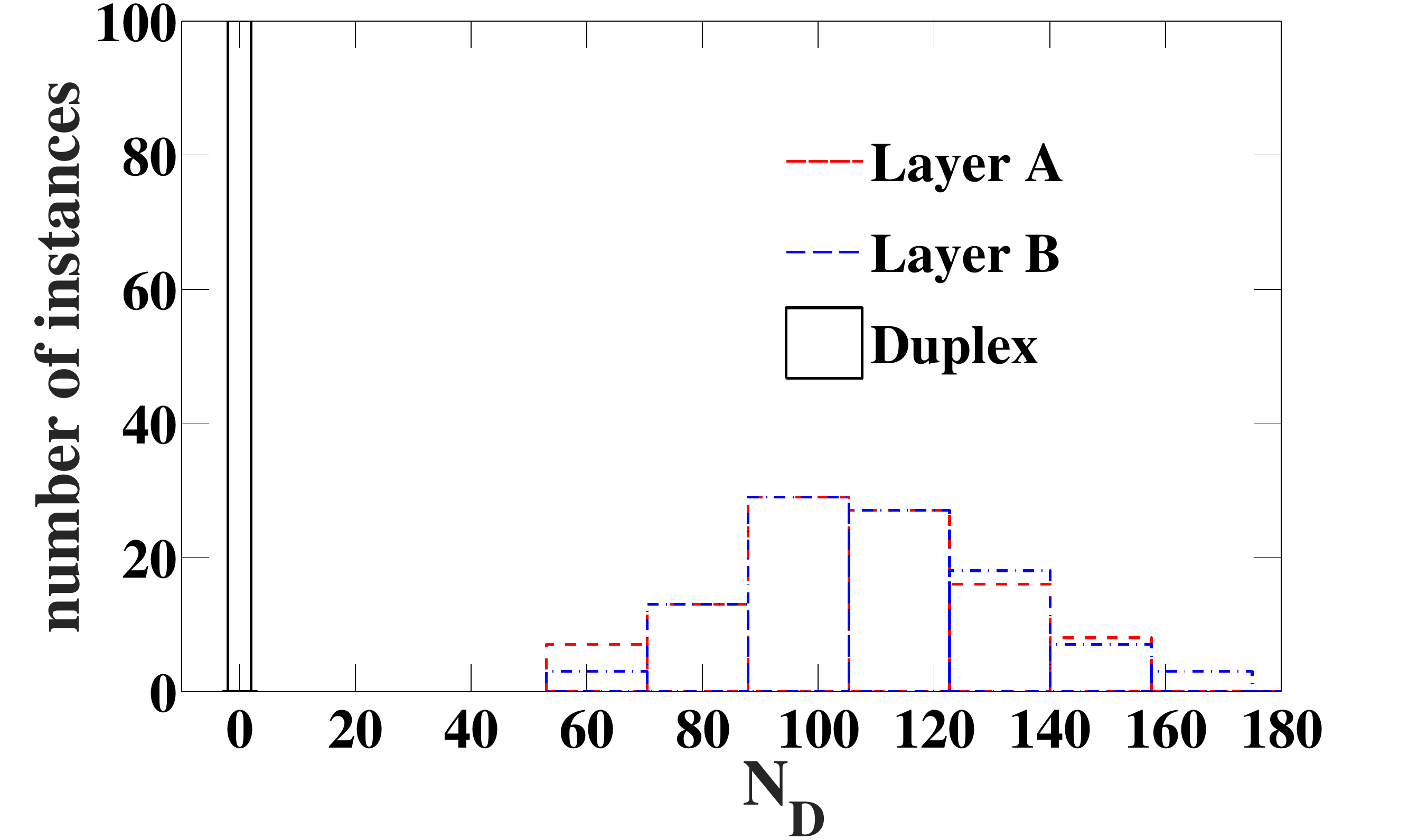}}
\end{center}
\caption{{\bf The fully controllable solution can be stable for the multiplex network also if it is not stable for the single layers taken in isolation.} Histogram of the number of networks that out of 100 realizations have $N_D$ driver nodes. The results obtained  for the control of a  duplex networks and its two layers are compared. The duplex networks are formed by two scale-free networks with $N=10^4$ and   $P^{A,in}(k)=P^{B,in}\propto k^{-\gamma}$ for $k>2$ and $P^{A,out}(k)=P^{B,out}(k)\propto k^{-\gamma}$ for $k>2$,  with $\gamma=2.3$, the networks have minimum in-degree equal to 2 and minimum out-degree equal to 3 and $P^{A,in}(2)=P^{B,in}(2)=0.3$.} 
\label{fig:asym}
\end{figure}

\paragraph*{Effect of degree correlations on the controllability of duplex networks --}
We consider a model of duplex network in which the replica nodes of the directed random graphs in the two layers have correlated degrees. In particular, we consider a case in which only the low in-degree nodes (nodes with in-degree equal to $0,1,2$) are correlated (replica nodes in different layers have same degree with probability $p$) and a case in which the in-degrees of the replica nodes are equal with probability $p$ independently of their value (see Supplementary Material for details). The controllability of the network is affected by these correlations as shown in Figure~\ref{fig:corr}. In fact, the number of driver nodes $n_D$ decreases as the level of correlation increases. In duplex networks with Poisson degree distribution, low-degree correlations modify both the position of the hybrid transition and the size of the discontinuity. Once the replica nodes with low in-degree are correlated, a further correlation of the remaining replica nodes does not substantially change the number of driver nodes. This result confirms that structural controllability is essentially determined by the control of low degree nodes \cite{PRL}.

\paragraph*{Stability of the fully controllable solution --}
A fully controllable solution, in which a single driver node is necessary to control the whole duplex network, exists if the minimum in-degree and the minimum out-degree are both greater than $1$ in both layers. This solution of the  cavity equations gives the correct solution to the maximum matching problem describing the controllability of multilayer networks only if no instabilities take place. 
The stability conditions are then found by imposing that the  Jacobian of the systems of equations derived by the cavity method has   all its eigenvalues  $\lambda_i$  of  modulus less than one, i.e. $|\lambda_i|<1$. 
In random duplex networks with the same degree distribution in the two layers, the fully controllable solution is stable (see Supplementary Material for the details of the derivation) if and only if  
\bea\label{conditions}
P_{\alpha}^{out}(2) < \frac{\avg{k^{\alpha}}_{in}\avg{k^{\alpha}}_{out}}{2\Avg{k^{\alpha}(k^{\alpha}-1)}_{in}},
\eea
for $\alpha=A,B$. On single networks it was instead recently found \cite{PRL} that the fully controllable  configuration  is only stable for 
\bea
P^{in}(2)<\frac{\avg{k}_{in}^2}{2\Avg{k(k-1)}_{out}},\ \ 
P^{out}(2)<\frac{\avg{k}^2_{in}}{2\Avg{k(k-1)}_{in}}.
\eea
This implies that for multiplex networks with asymmetric in-degree and out-degree distributions it might occur that  the fully controllable solution is stable in the multiplex network but unstable in the single networks taken in isolation (see Figure \ref{fig:asym} for the characterization of the controllability of a similar type of multiplex networks).
Therefore a multiplex structure can help to stabilize the fully controllable solution.

%\paragraph*{Conclusions --}
In conclusion, within the framework of structural controllability, we have considered the controllability properties of multiplex networks in which the nodes are either driver nodes in all the layers or they are not driver nodes in any layer.
Our results show that controlling multiplex networks is more demanding, in terms of number of driver nodes, than controlling networks composed of a single layer. In random duplex networks with Poisson degree distribution, it is possible to observe a hybrid phase transition with a discontinuity in the number of driver nodes as a function of the average degree, that is phenomenologically similar to the emergence of mutually connected components. Close to this phase transition the duplex network exhibits an increased fragility to random damage. The existence of correlations between the degrees of replica nodes in different layers, in particular between low-degree nodes, has the effect of reducing the number of driver nodes necessary to control duplex networks. Finally, multiplex structure of networks can stabilize the fully controllable solution also if this solution is not stable in the single layers that form the multiplex network.

\section*{METHODS}

\paragraph*{The BP equations --} The BP equations of this problem are derived using the cavity method \cite{Mezard,Cavity,Weigt}  as described for the case of a single network problem in \cite{Lenka,Altarelli,Zecchina,Liu,PRL}. The same approximation methods can be applied here, as long as the structure of the interconnected layers is locally tree-like both within the layers and across them. Under the decorrelation (replica-symmetric) assumption, the cavity fields (or messages) $\{h_{i\to j}^{\alpha}\}$ and $\{\hat{h}_{i\to j}^{\alpha}\}$, defined on the directed links between neighboring nodes $(i,\alpha)$ and $(j,\alpha)$ in the same layer $\alpha = A,B$  satisfy the zero-temperature limit of the BP equations, also known as Max-Sum equations,
\bea
\nonumber h_{i\to j}^{\alpha} &=&-\max\left [-1,\max_{k\in\partial^{+}i\setminus j} \hat{h}_{k\to i}^{\alpha}\right]\label{unoBP}\\
\nonumber \hat{h}^{A}_{i\to j} &=&-\max\left[\max_{k\in\partial_{-}^{A} i\setminus j}h_{k\to i}^{A}, -\max_{k\in\partial^{B}_{-}j} h_{k\to i}^{B} \right] \label{dueBP} \\
\hat{h}^{B}_{i\to j} &=&-\max\left[\max_{k\in\partial_{-}^{B} i\setminus j}h_{k\to i}^{B}, -\max_{k\in\partial^{A}_{-}j} h_{k\to i}^{A} \right] \label{treBP} \label{bp}
\eea
in which the fields are defined to take values in the discrete set $\{-1,0,1\}$ and here and in the following we use the convention that the maximum over a null set is equal to $-1$ (see Supplementary Material  for details).
In terms of these fields, the energy $E$ in Eq.\eqref{ene} becomes 
\bea\label{eneh}
\nonumber E&=& - \sum_{\alpha}\sum_{i=1}^N \max\left[-1, \max_{k\in\partial_{+}^{\alpha}i} \hat{h}_{k \to i}^{\alpha}\right] + \sum_{\alpha}\sum_{(i,j)}\max\left[0, h^{\alpha}_{i \to j}+\hat{h}^{\alpha}_{j \to i}\right]\nonumber \\
 &&-\sum_{i=1,N} \max\left[0, \max_{k\in\partial_{-}^Ai} h^A_{k \to i}+\max_{k\in\partial_{-}^Bi} h^B_{k \to i}\right].
\label{Energy}
\eea

\paragraph*{ BP equations over ensemble of networks-}
Let us consider the case of uncorrelated duplex networks in which the degree of the same node in different layers are uncorrelated and there is no overlap of the links. In each layer $\alpha=A,B$ we consider a maximally random network with in-degree distribution $P^{\alpha,in}(k)$ and out-degree distribution $P^{\alpha,out}(k)$.
At the ensemble level, each link of (the infinitely large) random network forming layer $\alpha$ has the same statistical properties, that we describe through distributions ${\cal P}_{\alpha}(h^{\alpha})$ and $\hat{{\cal P}}_{\alpha}(\hat{h}^{\alpha})$ of cavity fields that are defined on the support of Eqs. \ref{bp}, i.e.
\bea
{\cal P}_{\alpha}(h^{\alpha})&=&w_1^{\alpha}\delta(h^{\alpha}-1)+w_2^{\alpha}\delta(h^{\alpha}+1)+w_3^{\alpha}\delta(h^{\alpha}),\nonumber \\
\hat{{\cal P}}_{\alpha}(\hat{h}^{\alpha})&=&\hat{w}_1^{\alpha}\delta(\hat{h}^{\alpha}-1)+\hat{w}_2^{\alpha}\delta(\hat{h}^{\alpha}+1)+\hat{w}_3^{\alpha}\delta(\hat{h}^{\alpha}),
\eea
where $\alpha=A,B$  and where the probabilities $w_1^{\alpha},w_2^{\alpha},w_3^{\alpha}$ are normalized $w_1^{\alpha}+w_2^{\alpha}+w_3^{\alpha}=1$ as well as the probabilities $\hat{w}_1^{\alpha},\hat{w}_2^{\alpha},\hat{w}_3^{\alpha}$ that satisfy the equation $\hat{w}_1^{\alpha}+\hat{w}_2^{\alpha}+\hat{w}_3^{\alpha}=1$.
The cavity method at the network ensemble level is also known as density evolution method \cite{Mezard}.

It is useful to introduce the generating functions $G_0^{\alpha,in/out}(z),$ and $G_1^{\alpha,in/out}(z)$ of the multiplex network as $G_{0}^{\alpha,in/out}(z)=\sum_k P^{\alpha,in/out}(k)z^k,$
$G_1^{\alpha,in/out}(z)=\sum_k\frac{k}{\avg{k^{\alpha}}}P^{\alpha,in/out}(k)z^{k-1},$
with $\alpha=A,B$. In this way, we can derive recursive equations for the probabilities $\{w_i^{\alpha}\}_{i=1,2,3}$ and $\{\hat{w}_i^{\alpha}\}_{i=1,2,3}$, that are the analogous of the BP equations for an ensemble of uncorrelated duplex networks
\bea
w_1^{\alpha}&=&G_{1}^{\alpha,out}(\hat{w}_2^{\alpha}),\nonumber \\
w_2^{\alpha}&=&\left[1-G_{1}^{\alpha,out}(1-\hat{w}_1^{\alpha})\right],\nonumber \\
\hat{w}_1^{A}&=&G_{1}^{A,in}(w_2^A)\left[1-G_{0}^{B,in}(1-{w}_1^{B})\right],\nonumber \\ 
\hat{w}_2^{A}&=&\left[1-G_{1}^{A,in}(1-w_1^A)+G_1^{A,in}(1-w_1^A)G_0^{B,in}\left(w_2^B\right)\right],\nonumber \\
\hat{w}_1^{B}&=&G_{1}^{B,in}(w_2^B)\left[1-G_{0}^{A,in}(1-{w}_1^{A})\right],\nonumber \\ 
\hat{w}_2^{B}&=&\left[1-G_{1}^{B,in}(1-w_1^B)+G_1^{B,in}(1-w_1^B)G_0^{A,in}\left(w_2^A\right)\right],
\label{equncorr}
\eea
with $w_3^{\alpha}=1-w_1^{\alpha}-w_2^{\alpha},$ and 
$\hat{w}_3^{\alpha}=1-\hat{w}_1^{\alpha}-\hat{w}_2^{\alpha}$.
The energy $E$ and the entropy density $s$ of the matching problem can be also expressed in terms of the  $\{w_i^{\alpha}\}_{i=1,2,3}$ and $\{\hat{w}_i^{\alpha}\}_{i=1,2,3}$ (see Supplementary Material for details).

\paragraph*{Hybrid transition for Poisson duplex network--}
Here we consider the case  of two Poisson networks with the same in/out average degree. 
In other words, we consider the situation in which
$\Avg{k^{A,in}}=\Avg{k^{A,out}}=\Avg{k^{B,in}}=\Avg{k^{B,out}}=c$.
We notice that the BP equations  can be rewritten to form a closed subsystem of equations for $\hat{w}_1$ and $\hat{w}_2$ (see Supplementary Material for details), 
\begin{subequations}
\bea
\hat{w}_1&=&h_1(\hat{w}_1,\hat{w}_2)=e^{-ce^{-c\hat{w}_1}}\left[1-e^{-ce^{-c(1-\hat{w}_2)}}\right] 
\label{2w1hatsameDN}\\ 
\hat{w}_2&=&h_2(\hat{w}_1,\hat{w}_2)=\left[1-e^{-c e^{-c(1-\hat{w}_2)}}+e^{-ce^{-c(1-\hat{w}_2)}}e^{-ce^{-c
\hat{w}_1}}\right] \label{2w2hatsameDN}
\eea\label{hatsameDN}
\end{subequations}
from the solution of which the remaining quantities can be determined.
%
%\bea
%\hat{w}_1&=&e^{-ce^{-c\hat{w}_1}}\left[1-e^{-ce^{-c(1-\hat{w}_2)}}\right], 
%\label{w1hatsameDN}\\ 
%\hat{w}_2&=&\left[1-e^{-c e^{-c(1-\hat{w}_2)}}+e^{-ce^{-c(1-\hat{w}_2)}}e^{-ce^{-c
%\hat{w}_1}}\right], \label{w2hatsameDN}\\
%w_1&=&e^{-c(1-\hat{w}_2)},\nonumber \\
%w_2&=&\left[1-e^{-c\hat{w}_1}\right],\nonumber \\
%w_3&=&1-w_1-w_2,\nonumber \\
%\hat{w}_3&=&1-\hat{w}_1-\hat{w}_2.\nonumber
%\eea

The value $c^{\star}$ of the average degree $c$ at which the discontinuity in the number of driver nodes $n_D$ observed is found by imposing that Eqs.~\eqref{hatsameDN} are satisfied together with the condition
\bea
\left|J\right|=0,
\label{J}
\eea
with $J$ indicating the Jacobian of the system of equations (\ref{hatsameDN}).
Imposing that Eqs.~\eqref{hatsameDN} and condition \eqref{J} are simultaneously satisfied, the solution $c^{\star}=3.222326106\ldots$ is found.
For $c<c^{\star}$ we observe that $w_3=\hat{w}_3=0$. At $c^{\star}$ we observe a discontinuity in both $w_3$ and $\hat{w}_3$, but for $c>c^{\star}$ the functions $h_1(\hat{w}_1,\hat{w}_2)$ and $h_2(\hat{w}_1,\hat{w}_2)$ are analytic, and analyzing Eqs. $(\ref{2w1hatsameDN})-(\ref{2w2hatsameDN})$ we obtain the behaviour of the order parameters $w_3$ and $\hat{w}_3$ for $c>c^{\star}$
\bea
w_3-w_3^{\star}&\propto &(c-c^{\star})^{1/2}\nonumber \\
\hat{w}_3-\hat{w}_3^{\star}&\propto &(c-c^{\star})^{1/2},
\eea
showing that the transition is hybrid.\\
\section*{Acknowledgements}
LD aknowledges the European Research Council for grant n. 267915.
GM aknowledges the European Project MIMOmics.
\section*{Author Contributions}
GM, LD and GB designed the study, developed the methodology, performed the analysis and wrote the manuscript.
The authors declare no competing financial interests.
\newpage
\section*{Supplemental Material "Control of Multilayer Networks"}
\subsection{Introduction}
\noindent This Supplemental Material is structured as follows.\\
In Sec. II we define the problem of structural controllability of multiplex networks, focusing on the case of a duplex network. Moreover we define the driver nodes, as the set of nodes that, if  stimulated by an external signal, can drive the dynamical state of the network to any desired configuration.\\
In Sec. III we map the problem of structural controllability of a duplex network to a Maximum Matching Problem, and we derive the Belief Propagation (BP) equations determining the driver nodes, and their zero-temperature limit known as Max-Sum equations.\\
In Sec IV we consider the controllability of uncorrelated duplex networks, characterizing the BP equations valid for this problem,  the stability conditions  for the solutions of the BP equations, and the entropy of the solutions. Moreover, we consider  duplex networks formed  by two  Poisson layers and we characterize their hybrid phase transition. Finally we consider the controllability of   duplex networks formed by layers with power law in-degree and out-degree distributions.\\
In Sec. V we consider ensembles of duplex networks in which the in-degrees of replica nodes are correlated and we derive the BP equations assuming either that only the low in-degrees  of replica nodes are correlated or that all the in-degrees of replica nodes are correlated.

\subsection{The structural controllability of a multiplex network}

We consider a multiplex network in which  every node $i=1,2,\ldots, N$ has a replica node in each layer and every layer is formed by a directed networks between the corresponding replica nodes \cite{PhysReports}. We assume that each replica node can have a different dynamical state and can send different signals in the different networks (each layers is characterized by a different dynamical process). In this case the controllability of the multiplex network can be treated by control theory methods  used for the single layers taken in isolation \cite{slotine,Lin74,Liu,PRL}.  Nevertheless here we will consider an additional constraint on the number of driver nodes. In fact we impose that corresponding replica nodes are either driver nodes in all layers or they are not driver nodes in any layer.

We  consider for simplicity a duplex, i.e a multiplex formed by two layers where each layer is formed by a directed network.
We call the two layers layer $\alpha=A,B$. 
We consider a  linear dynamical system  determining the network dynamics
\bea
\frac{d{\bf X}(t)}{dt}={\cal G}{\bf X}(t)+{\cal K}{\bf u}(t),
\label{dyn}
\eea
in which the vector ${\bf X}(t)$ describes the dynamical state of each replica node  in the duplex, and has $2N$ elements. The first set of $N$   elements represents the dynamical state $x_i^{A}$ of node $i$ in layer A  (i.e. $X_{i}=x_i^{A}$ for $i=1,2\ldots,N$), while the elements $X_{N+i}$ represent the dynamical state of the node $i$ in layer B, and are given by $X_{N+i}=x_i^{B}$ for $i=1,2,\ldots, N$.
The matrix ${\cal G}$ is a $2N\times 2N$ (asymmetric) matrix and the matrix ${\cal K}$ is a $2N\times M$ matrix. The matrices ${\cal G}$ and ${\cal K}$  have  the following block structure
\bea
{\cal G}=\left(\begin{array}{cccc} g^{A}&0 \\ 0& g^{B}  \end{array}\right), \ \ \ {\cal K}=\left(\begin{array}{cc}K^{A}& 0 \\ 0& K^{B}\end{array}\right),
\label{bB}
\eea 
where $g^{\alpha}$ with $\alpha=A,B$  are the $N\times N$ matrices describing the directed weighted interactions within each of the  networks in the two layers and    ${ K}^{\alpha}$ are the $N\times M^{\alpha}$ matrices  describing the interaction between the nodes of the network $\alpha$ and the $M^{\alpha}\leq N$ external signals for  layer $\alpha$. The external signals are indicated by the vector ${\bf u}(t)$ of elements $u_{\gamma}$ and $\gamma=1,2\ldots M=M^{A}+M^{B}$.

Given block structure of both matrix ${\cal G}$ and ${\cal K}$ described in Eq. $(\ref{bB})$, the problem of duplex network controllability defined by Eq. $(\ref{dyn})$, can be exactly recast into the problem of controllability of the single layers that form the duplex network.

Here we adopt the framework of {\em structural controllability} \cite{Lin74} aimed at characterizing if a given duplex network is controllable when the non-zero matrix elements of ${\cal G}$ and ${\cal K}$  given by  Eq. $(\ref{bB})$ are free parameters.
A duplex networks in which the linear dynamics described by the Eqs. $(\ref{dyn})$ and $(\ref{bB})$ take place, is structurally controllable if both layers $\alpha=A,B$ are structurally controllable. 
%For this reasons, if the matrices ${\cal G}$ and ${\cal K}$ are given, in principle it can be possible to use the Kalman's condition for checking the multiplex network controllability. Nevertheless, determining  the minimum number of driver nodes, (i.e. the minimum number of controlled nodes that do not share an input) using Kalman's condition, can be computationally too costly for been implementable, in particular for large networks.
% Moreover it can happen frequently that the weighted entries of the matrices ${\cal G}$ and ${\cal K}$ are know only approximately. In this case small perturbations to the entries of the matrices  ${\cal G}$ and ${\cal K}$  could affect the result found using Kalman's condition for controllability.
%The structural controllability of a single network has been introduced in \cite{Lin74}, and then mapped to the Maximum Matching problem by Liu et al. in \cite{Liu}.
%Here we will describe how the framework of structural controllability applied to  duplex  of networks by a dynamics defined by Eq. $(\ref{dyn})$.

Each layer $\alpha$ is structurally controllable if for any choice of the free parameters in ${g^{\alpha}}$ and ${ K^{\alpha}}$, except for a variety of zero Lebesgue measure in the parameter space,  the Kalman's condition is fulfilled \cite{Lin74}. 
Since structural controllability only distinguishes between zero and non-zero entries of the matrices $g^{\alpha}$ and ${K^{\alpha}}$, a given directed network in layer $\alpha$ is structurally controllable if it is possible to determine the input nodes (i.e. the position of the non-zero entries of the matrix ${K^{\alpha}}$) in a way to control the dynamics described by any realization of the matrix ${g^{\alpha}}$ with the same non-zero elements, except for atypical realizations of zero measure.
In practice, a single network can be structurally controlled by identifying a minimum number of {\em driver nodes}, that are controlled nodes which do not share input vertices in both layers. 
In their seminal paper \cite{Liu}, Liu and coworkers showed that on single networks this control theoretic problem can be reduced to a well-known optimization problem: their Minimum Input Theorem states that the minimum set of driver nodes that guarantees the full structural controllability of a network is the set of unmatched nodes in a maximum  matching of the same directed network.
Their result for a single network remains valid for the duplex network described by Eqs. $(\ref{dyn})-(\ref{bB})$. 
Therefore the structural controllability of  duplex networks,  in the absence of further constraints can be mapped to a Maximum Matching problem defined on the single layers of  the duplex networks.
Here nevertheless, we consider a further constraint to be imposed on the driver nodes, which enforce a new type of dependence between the layers of the duplex.
In particular we impose that the replica nodes $(i,\alpha)$ with  $\alpha=A,B$ and a given  index $i$, are either both driver nodes or neither is a driver node. This implies that these two replica nodes  are either both linked to independent and external signals or none of them is connected to external signals.

\subsection{The Maximum Matching Problem for the Controllability of Duplex Networks}
\subsubsection{Mapping duplex controllability into a constrained Maximum Matching Problem}
\label{MMP}

In order to build an algorithm able to find the driver nodes of a  duplex network  we consider the variables $s_{ij}^{\alpha}=1,0$ indicating respectively if the directed link from node  $(i,\alpha)$ to node $(j,\alpha)$ in layer $\alpha=A,B$ is matched or not.
In the two layers of the duplex network we want to have a matching, i.e. the following constraints must always be satisfied for $\alpha=A,B$,
\begin{subequations}
	\bea
	\sum_{j\in \partial_{+}^{\alpha}i}s_{ij}^{\alpha}&\leq& 1, \\
	\sum_{j\in \partial_{-}^{\alpha}i}s_{ji}^{\alpha}&\leq& 1,
	\eea \label{uno}
\end{subequations}
where here and in the following we indicate with $\partial_+^{\alpha}$ the set of nodes $j$ that are pointed by node $i$ in layer $\alpha$ and with $\partial_-^{\alpha}i$ the set of nodes $j$ pointing to node $i$ in layer $\alpha$.
In addition we impose that  the driver nodes in the two networks are replica nodes, i.e. in the matching problem either two replica nodes are both matched or both unmatched.  Therefore the variable $s_{ij}^{\alpha}$ satisfy the following additional constraints
\bea
\sum_{i\in \partial_{-}^{A}j}s_{ji}^{A}=\sum_{i\in \partial_{-}^{B}j}s_{ji}^{B}.
\label{due}
\eea
Finally we need to minimize the number of driver nodes in the multiplex network. 
Therefore we  minimize the  energy $E$ of the problem  given by 
\bea
E&=&\sum_{\alpha}\sum_j\left(1-\sum_{i\in \partial_{-}^{\alpha}j}s_{ij}^{\alpha}\right)=\sum_{\alpha}\sum_{i}E_i^{\alpha},
\label{E}
\eea
with
\bea
E_i^{\alpha}=1-\sum_{j\in \partial_+^{\alpha}i}s_{ij}^{\alpha}. 
\eea
The energy $E$ is given by the number $N_D$ of driver replica nodes in the duplex network by 
\bea 
E=N_D=Nn_D.
\eea
\subsubsection{Derivation of the BP equations at finite inverse temperature $\beta$}
We consider here the Maximum Matching Problem defined in Sec \ref{MMP}. The goal is to find the configuration of the variables $\{s_{ij}^{\alpha}\}$ associated to every directed edge $i\to j$ in layer $\alpha$, such that the energy $E$ given by the number of driver replica nodes in the duplex network is minimized provided that the conditions given by Eqs. $(\ref{uno})$, $(\ref{due})$ are satisfied. 
Introducing as an auxiliary variable the ``inverse temperature" $\beta$ we cast this problem into a statistical mechanics problem  where our first aim  is finding the distribution $P(\{s_{ij}\})$, parametrized by the {\it inverse temperature} $\beta$, and  given by 
\bea
P(\{s_{ij}\})&=&\frac{e^{-\beta E}}{Z}\prod_{i=1}^N\left\{\prod_{\alpha}\left[\theta\left(1-\sum_{j\in \partial_{+}i}s_{ij}^{\alpha}\right)\theta\left(1-\sum_{j\in \partial_{-}i}s_{ji}^{\alpha}\right) \right]\delta\left(\sum_{i\in \partial_{-}^{A}j}s_{ij}^{A},\sum_{i\in \partial_{-}^{B}j}s_{ij}^{B}\right)\right\},
\label{psij}
\eea
where $\theta(x)=1$ for $x\geq 0$ and $\theta(x)=0$ for $x<0$ , $\delta (x)$ is the Kronecker delta, and where $Z$ is the normalization constant, that corresponds to the partition function of the statistical mechanics problem. Subsequently, we plan to perform the limit $\beta\to \infty$ in order to characterize the optimal (i.e. the maximum-sized) matching in the network satisfying Eqs. $(\ref{uno})$, $(\ref{due})$.
The free-energy of  the problem $F(\beta)$ is defined as 
\bea
\beta F(\beta)=-\ln Z,
\eea
and the energy $E$ is therefore given by 
\bea
E=\frac{\partial [\beta F(\beta)]}{\partial \beta}.
\label{der}
\eea

The distribution $P(\{s_{ij}\})$ on a locally tree-like network can be (approximately) estimated by the cavity method in the replica symmetric assumption (i.e. by deriving Belief Propagation equations) \cite{Liu,PRL,Lenka,Altarelli,Mezard,Cavity, Zecchina,Weigt}.
In this respect,  in each layer $\alpha$ of the duplex network, we define two probability marginals on each directed link, one going in the same direction of the link  $P_{i\to j}^{\alpha}(s_{ij})$ and one in the opposite direction $\hat{P}_{i\to j}^{\alpha}(s_{ji})$. The BP equations for these quantities are 
\begin{subequations}
	\bea
	P_{i\to j}^{\alpha}(s_{ij})&=&\frac{1}{{\cal D}_{i \to j}^{\alpha}} \sum_{\{s_{ ik}^{\alpha} \} |k \in \partial_{+}^{\alpha}i\setminus j}  \left\{\theta\left(1-\sum_{k\in \partial_{+}^{\alpha}i}s_{ik}^{\alpha}\right)\exp{ \left [-\beta \left(1- \sum_{k\in \partial_{+}^{\alpha}i}s_{ik}^{\alpha} \right) \right]}\prod_{k\in \partial_+^{\alpha}i\setminus j}\hat{P}_{k\to i}^{\alpha}(s_{ik}^{\alpha})\right\}, \\
	\hat{P}_{i\to j}^{A}(s_{ji}^{A})&=&\frac{1}{{\hat{\cal D}}_{i \to j}^{A}}   \sum_{\{s_{ ki }^{A}\} \setminus s_{ji}^{A}, k \in \partial_{-}^{A}i}\left\{\theta\left(1-\sum_{k\in \partial_{-}^{A}i}s_{ki}^{A}\right) \sum_{\{s_{ ki }^{B}\} | k \in \partial_{-}^{B}i}\left[\theta\left(1-\sum_{k\in \partial_{-}^{B}i}s_{ki}^{B}\right)\delta\left(\sum_{i\in \partial_{-}^{A}j}s_{ij}^{A},\sum_{i\in \partial_{-}^{B}j}s_{ij}^{B}\right)\right. \right.\nonumber \\
	&&\times\left.\left. \prod_{k\in \partial_-^{A}i\setminus j}{P}_{k\to i}^{A}(s_{ki}^{A}) \prod_{k\in \partial_-^{B}i}{P}_{k\to i}^{B}(s_{ki}^{B})\right]\right\}, \\
	\hat{P}_{i\to j}^{B}(s_{ji}^{B})&=&\frac{1}{{\hat{\cal D}}_{i \to j}^{B}}   \sum_{\{s_{ ki }^{B}\} | k \in \partial_{-}^{B}\setminus j}\left\{\theta\left(1-\sum_{k\in \partial_{-}^{B}i}s_{ki}^{B}\right) \sum_{\{s_{ ki }^{A}\} | k \in \partial_{-}^{A}i}\left[\theta\left(1-\sum_{k\in \partial_{-}^{A}i}s_{ki}^{A}\right)\delta\left(\sum_{i\in \partial_{-}^{A}j}s_{ij}^{A},\sum_{i\in \partial_{-}^{B}j}s_{ij}^{B}\right)\right. \right. \nonumber \\
	&&\times \left. \left. \prod_{k\in \partial_-^{B}i\setminus j}{P}_{k\to i}^{B}(s_{ki}^{B}) \prod_{k\in \partial_-^{A}i}{P}_{k\to i}^{A}(s_{ki}^{A})\right]\right\}
	\eea  \label{BPe}
\end{subequations}
where ${\cal D}_{i \to j}^{\alpha}$ and ${\hat{\cal D}}_{i \to j}^{\alpha}$ are normalization constants.
The probability marginals $\{P_{i\to j}^{\alpha}(s_{ij}^{\alpha})$,  $\hat{P}_{i\to j}^{\alpha}(s_{ji}^{\alpha})\}$ can be parametrized by the cavity fields $h_{i\to j}^{\alpha}$ and $\hat{h}_{i\to j}^{\alpha}$ defined by 
\bea
&\begin{array}{lr}
	P_{i\to j}^{\alpha}(s_{ij}^{\alpha})=\frac{\exp\left[{\beta h_{i\to j}^{\alpha}s_{ij}^{\alpha}}\right]}{1+\exp\left[{\beta h_{i\to j}^{\alpha}}\right]}\ \ \ \ \ \ \  \hat{P}_{i\to j}^{\alpha}(s_{ji}^{\alpha})=\frac{\exp\left[{\beta\hat{ h}_{i\to j}^{\alpha}s_{ji}^{\alpha}}\right]}{1+\exp\left[{\beta \hat{h}_{i\to j}^{\alpha}}\right]}.
	& \end{array}
\eea  
%\end{document}
In terms of the cavity fields (or messages), Eqs.~(\ref{BPe}) reduce to the following set of finite temperature BP equations,
\begin{subequations}
	\bea
	h_{i\to j}^{\alpha}&=&-\frac{1}{\beta} \log \left(e^{-\beta}+\sum_{k\in\partial_{+}^{\alpha}i\setminus j}e^{\beta \hat{h}_{k\to i}^{\alpha}}\right), \\
	\hat{h}_{i\to j}^{A}&=&-\frac{1}{\beta} \log \left(\frac{1}{\sum_{k\in\partial_{-}^B i}e^{\beta h_{k\to i}^B}}+\sum_{k\in\partial_{-}^{A}i\setminus j}e^{\beta {h}^A_{k\to i}}\right), \\
	\hat{h}_{i\to j}^{B}&=&-\frac{1}{\beta} \log \left(\frac{1}{\sum_{k\in\partial_{-}^A i}e^{\beta h^A_{k\to j}}}+\sum_{k\in\partial_{-}^{B}i\setminus j}e^{\beta {h}^B_{k\to i}}\right),
	\eea
	\label{BPbeta}
\end{subequations}

The free energy $F$ and the energy $E=\frac{\partial \beta F}{\partial \beta}$ of the  model are given respectively by 

\bea
-\beta F&=&\sum_{\alpha}\sum_{i=1}^N \left[\ln\left(e^{-\beta}+\sum_{k\in \partial_{+}^{\alpha}i}e^{\beta \hat{h}^{\alpha}_{k\to i}}\right)\right]+\sum_{i=1,N}
\ln\left(1+\sum_{k\in \partial_{-}^{A}i}e^{\beta h_{k\to i}^A}\sum_{k'\in \partial_{-}^{B}i}e^{\beta h_{k'\to i}^{B}}\right)
\nonumber \\
&&-\sum_{\alpha}\sum_{<i,j>_{\alpha}}\ln \left(1+e^{\beta(h_{i\to j}^{\alpha}+\hat{h}_{j\to i}^{\alpha})}\right),
\eea
and by 
\bea
E&=&\sum_{\alpha}\sum_{i=1}^N \left[\frac{e^{-\beta}-\sum_{k\in \partial_{+}^{\alpha}i}\hat{h}^{\alpha}_{k\to i}e^{\beta \hat{h}^{\alpha}_{k\to i}}}{e^{-\beta}+\sum_{k\in \partial_{+}^{\alpha}i}e^{\beta \hat{h}^{\alpha}_{k\to i}}}\right]-\sum_{i=1}^N
\frac{\sum_{k\in \partial_{-}^{A}i}h_{k\to i}^Ae^{\beta h_{k\to i}^A}\sum_{k'\in \partial_{-}^{B}i}e^{\beta h_{k'\to i}^{B}}}{1+\sum_{k\in \partial_{-}^{A}i}e^{\beta h_{k\to i}^A}\sum_{k'\in \partial_{-}^{B}i} e^{\beta h_{k'\to i}^{B}}}\nonumber \\
&-&\sum_i \frac{\sum_{k\in \partial_{-}^{A}i}e^{\beta h_{k\to i}^A}\sum_{k'\in \partial_{-}^{B}i}h_{k'\to i}^{B}e^{\beta h_{k'\to i}^{B}}}{1+\sum_{k\in \partial_{-}^{A}i}e^{\beta h_{k\to i}^A}\sum_{k'\in \partial_{-}^{B}i} e^{\beta h_{k'\to i}^{B}}}
+\sum_{\alpha}\sum_{<i,j>_{\alpha}}\frac{(h_{i\to j}^{\alpha}+\hat{h}_{j\to i}^{\alpha})e^{\beta(h_{i\to j}^{\alpha}+\hat{h}_{j\to i}^{\alpha})}}{ 1+e^{\beta(h_{i\to j}^{\alpha}+\hat{h}_{j\to i}^{\alpha})}}.
\eea

\subsubsection{BP Equations for $\beta \to \infty$}
The BP equations in the limit $\beta \to \infty$ are derived from the Eqs. $(\ref{BPbeta})$.  
In the limit $\beta\to \infty$ the  solution is expressed in terms of the fields $h_{i\to j}^{\alpha}$ or $\hat{h}_{i\to j}^{\alpha}$ sent from a node $(i,\alpha)$ to the linked node $(j,\alpha)$ in layer $\alpha=A,B$. The cavity fields have a simple interpretation as messages between neighboring replica nodes \cite{Lenka}: $h_{i\to j}^{\alpha}=\hat{h}_{i\to j}^{\alpha}=1$  means {\em ``match me''}, $h_{i\to j}^{\alpha}=\hat{h}_{i\to j}^{\alpha}=-1$ means {\em ``do not match me''}, and $h_{i\to j}^{\alpha}=\hat{h}_{i\to j}^{\alpha}=0$ means {\em ``do what you want''}. 
The zero-temperature BP (or Max-Sum) equations determining the values of these fields in the limit $\beta\to \infty$ are are given by
\begin{subequations} 
	\bea
	h_{i\to j}^{\alpha} &=&-\max\left [-1,\max_{k\in\partial^{+}i\setminus j} \hat{h}_{k\to i}^{\alpha}\right]\label{unoBP}\\
	\hat{h}^{A}_{i\to j} &=&-\max\left[\max_{k\in\partial_{-}^{A} i\setminus j}h_{k\to i}^{A}, -\max_{k\in\partial^{B}_{-}j} h_{k\to i}^{B} \right] \label{dueBP}\\
	\hat{h}^{B}_{i\to j} &=&-\max\left[\max_{k\in\partial_{-}^{B} i\setminus j}h_{k\to i}^{B}, -\max_{k\in\partial^{A}_{-}j} h_{k\to i}^{A} \right] \label{treBP}
	\eea\label{maxsum}
\end{subequations}
in which the fields are defined to take values in the discrete set $\{1,0,-1\}$ and we defined the maximum over a null set equal to $-1$. 
It follows that for $k^{{B},in}_i=0$ we have $\hat{h}^{A}_{i\to j} =-1$ and for $k^{{A},in}_i=0$ we have $\hat{h}^{B}_{i\to j} =-1$.

The energy $E$ can also be expressed in terms of these fields and is given by 
\bea
\nonumber E&=& - \sum_{\alpha}\sum_{i=1}^N \max\left[-1, \max_{k\in\partial_{+}^{\alpha}i} \hat{h}_{k \to i}^{\alpha}\right]\\
& & + \sum_{\alpha}\sum_{<i,j>}\max\left[0, h^{\alpha}_{i \to j}+\hat{h}^{\alpha}_{j \to i}\right]\nonumber \\
&&-\sum_{i=1,N} \max\left[0, \max_{k\in\partial_{-}^Ai} h^A_{k \to i}+\max_{k\in\partial_{-}^Bi} h^B_{k \to i}\right],
\label{EnergyBP}
\eea
where $<i,j>$ indicates pair of nodes that are nearest neighbors in the network  and where we take the maximum over a null set equal to -1.

\subsection{Controllability of uncorrelated multiplex networks with given in-degree and out-degree distribution}
\subsubsection{Cavity equations for an uncorrelated multiplex network ensemble}
Let us consider the case of uncorrelated duplex networks in which the degree of the same node in different layers are uncorrelated and there is no overlap of the links. In each layer $\alpha=A,B$ we consider a maximally random network with in-degree distribution $P^{\alpha,in}(k)$ and out-degree distribution $P^{\alpha,out}(k)$.
At the ensemble level, each link of (the infinitely large) random network forming layer $\alpha$ has the same statistical properties, that we describe through distributions ${\cal P}_{\alpha}(h^{\alpha})$ and $\hat{{\cal P}}_{\alpha}(\hat{h}^{\alpha})$ of cavity fields that are defined on the support of Eqs.\ref{maxsum}, i.e.
\bea
{\cal P}_{\alpha}(h^{\alpha})&=&w_1^{\alpha}\delta(h^{\alpha}-1)+w_2^{\alpha}\delta(h^{\alpha}+1)+w_3^{\alpha}\delta(h^{\alpha}),\nonumber \\
\hat{{\cal P}}_{\alpha}(\hat{h}^{\alpha})&=&\hat{w}_1^{\alpha}\delta(\hat{h}^{\alpha}-1)+\hat{w}_2^{\alpha}\delta(\hat{h}^{\alpha}+1)+\hat{w}_3^{\alpha}\delta(\hat{h}^{\alpha}),
\eea
where $\alpha=A,B$  and where the probabilities $w_1^{\alpha},w_2^{\alpha},w_3^{\alpha}$ are normalized $w_1^{\alpha}+w_2^{\alpha}+w_3^{\alpha}=1$ as well as the probabilities $\hat{w}_1^{\alpha},\hat{w}_2^{\alpha},\hat{w}_3^{\alpha}$ that satisfy the equation $\hat{w}_1^{\alpha}+\hat{w}_2^{\alpha}+\hat{w}_3^{\alpha}=1$.
The cavity method at the network ensemble level is also known as density evolution method \cite{Mezard}.

It is useful to introduce the generating functions $G_0^{\alpha,in/out}(z),$ and $G_1^{\alpha,in/out}(z)$ of the multiplex network as 
\bea
G_{0}^{\alpha,in}(z)&=&\sum_k P^{\alpha,in}(k)z^k,\nonumber \\
G_1^{\alpha,in}(z)&=&\sum_k\frac{k}{\avg{k^{\alpha}}}P^{\alpha,in}(k)z^{k-1},\nonumber \\
G_{0}^{\alpha,out}(z)&=&\sum_k P^{\alpha,out}(k)z^k,\nonumber \\
G_1^{\alpha,out}(z)&=&\sum_k\frac{k}{\avg{k^{\alpha}}}P^{\alpha,out}(k)z^{k-1},
\label{generating}
\eea
with $\alpha=A,B$. In this way, we can derive recursive equations for the probabilities $\{w_i^{\alpha}\}_{i=1,2,3}$ and $\{\hat{w}_i^{\alpha}\}_{i=1,2,3}$, that are the analogous of Eqs.~\ref{maxsum} for an ensemble of uncorrelated duplex networks
\bea
w_1^{\alpha}&=&G_{1}^{\alpha,out}(\hat{w}_2^{\alpha}),\nonumber \\
w_2^{\alpha}&=&\left[1-G_{1}^{\alpha,out}(1-\hat{w}_1^{\alpha})\right],\nonumber \\
w_3^{\alpha}&=&1-w_1^{\alpha}-w_2^{\alpha},\nonumber \\
\hat{w}_3^{\alpha}&=&1-\hat{w}_1^{\alpha}-\hat{w}_2^{\alpha},\nonumber\\
\hat{w}_1^{A}&=&G_{1}^{A,in}(w_2^A)\left[1-G_{0}^{B,in}(1-{w}_1^{B})\right],\nonumber \\ 
\hat{w}_2^{A}&=&\left[1-G_{1}^{A,in}(1-w_1^A)+G_1^{A,in}(1-w_1^A)G_0^{B,in}\left(w_2^B\right)\right],\nonumber \\
\hat{w}_1^{B}&=&G_{1}^{B,in}(w_2^B)\left[1-G_{0}^{A,in}(1-{w}_1^{A})\right],\nonumber \\ 
\hat{w}_2^{B}&=&\left[1-G_{1}^{B,in}(1-w_1^B)+G_1^{B,in}(1-w_1^B)G_0^{A,in}\left(w_2^A\right)\right].
\label{equncorr}
\eea

The energy $E$ of the matching problem can be also expressed in terms of the  $\{w_i^{\alpha}\}_{i=1,2,3}$ and $\{\hat{w}_i^{\alpha}\}_{i=1,2,3}$ giving
\bea
E&=&\sum_{\alpha}\left\{G_0^{\alpha,out}\left(\hat{w}_2^{\alpha}\right)-\left[1-G_{0}^{\alpha,out}(1-\hat{w}_1^{\alpha})\right]\right\}
-\left\{[1-G_{0}^{A,in}(1-w_1^A)][1-G_{0}^{B,in}(w_2^B)]\right.\nonumber \\
&&\left.+[1-G_{0}^{B,in}(1-w_1^B)][1-G_0^{A,in}(w_2^A)]\right\}+\sum_{\alpha}{\avg{k^{\alpha}}_{in}}\left[\hat{w}_{1}^{\alpha}(1-w_2^{\alpha})+w_{1}^{\alpha}(1-\hat{w}_2^{\alpha})\right].
\eea

\subsubsection{Stability condition}
\label{stabcondduplex}
The Eqs.\ref{equncorr} might have multiple solutions. In order to evaluate the stability of these solutions, using a method already used in the context of single networks \cite{PRL,Lenka}
here we compute  the Jacobian of the system of Eqs.~\eqref{equncorr} and impose that all its eigenvalues have modulus less than one. We avoid to consider $w_3^{\alpha}$ and $\hat{w}_3^{\alpha}$ because they influence only the number of null eigenvalues (4 eigenvalues upon 12). The $12\times12$ Jacobian matrix becomes $8\times8$ and it can be decomposed in four $4\times4$ blocks  
\bea
J=\left(\begin{array}{cc}H_{11}& H_{1,2}\nonumber \\
	H_{2,1}& H_{2,2}\end{array}\right).
\eea
with 
\bea
H_{11}=\left(\begin{array}{cccc}0&0&0&G_{2}^{A,out}(\hat{w}^{A}_2)\\
	0&0&G_{2}^{A,out}(1-\hat{w}^{A}_1)&0\\
	0& G_{2}^{A,in}(w^{A}_2)(1-G_{0}^{B,in}(1-w_1^{B})) & 0&0\\
	G_{2}^{A,in}(1-w_1^{A})(1-G_{0}^{B,in}(w_2^{B}))&0&0&0\\ \end{array}
\right),
\eea
\bea
H_{2,1}=\left(\begin{array}{cccc}0&0&0&0\\
	0&0&0&0\\
	G_{1}^{B,in}(w^{B}_2)\Avg{k}_{A,in}G_{1}^{A,in}(1-w^{A}_1)  &0&0&0\\
	0&G_{1}^{B,in}(1-w^{B}_1)\Avg{k}_{A,in}G_{1}^{A,in}(w^{A}_2) &0&0\end{array}\right),
\eea
\bea
H_{1,2}=\left(\begin{array}{cccc} 0&0&0&0\\
	0&0&0&0\\
	G_{1}^{A,in}(w^{A}_2)\Avg{k}_{B,in}G_{1}^{B,in}(1-w^{B}_1)  &0&0&0\\
	0&G_{1}^{A,in}(1-w^{A}_1)\Avg{k}_{B,in}G_{1}^{B,in}(w^{B}_2) &0&0
\end{array}\right),
\eea
and 
\bea
H_{2,2}=\left(\begin{array}{cccc}0&0&0&G_{2}^{B,out}(\hat{w}^{B}_2)\\
	0&0&G_{2}^{B,out}(1-\hat{w}^{B}_1)&0\\
	0& G_{2}^{B,in}(w^{B}_2)(1-G_{0}^{A,in}(1-w_1^{A})) &0& 0\\
	G_{2}^{B,in}(1-w_1^{B})(1-G_{0}^{A,in}(w_2^{A}))&0&0&0\\
\end{array}\right).
\eea
Here the  generating functions $G_0^{\alpha,in/out}$ and $G_1^{\alpha,in/out}$ are given by  Eqs. $(\ref{generating})$ and  the generating functions $G_{2}^{\alpha,in}(x)$ and $G_{2}^{\alpha,out}(x)$ are defined as 
\bea
G_{2}^{\alpha,in}(z)&=&\sum_{k}\frac{k(k-1)}{\avg{k^{\alpha}}_{in}}P_{\alpha}^{in}(k)z^{k-2}\nonumber\\
G_{2}^{\alpha,out}(z)&=&\sum_{k}\frac{k(k-1)}{\avg{k^{\alpha}}_{out}}P_{\alpha}^{out}(k)z^{k-2}.
\eea
Of particular interest is the characterization of the stability of the  solution $w^{\alpha}_1=\hat{w}^{\alpha}_1=w^{\alpha}_2=\hat{w}^{\alpha}_2=0$ and $w_3^{\alpha}=\hat{w}^{\alpha}_3=1$,  corresponding to the full controllability of the network, a configuration with $E=N_D=0$.
This solution emerges for $P_{\alpha}^{in}(1)=P_{\alpha}^{out}(1)=0$ for $\alpha=A,B$. Therefore if the minimum in-degree and the minimum out-degree are both greater than one, the analysis at the ensemble level is consistent with the full controllability of the network. Nevertheless this solution might be not stable. By analyzing the   Jacobian  $J$ for $w^{\alpha}_1=\hat{w}^{\alpha}_1=w^{\alpha}_2=\hat{w}^{\alpha}_2=0$ and $w_3^{\alpha}=\hat{w}^{\alpha}_3=1$, we can determine under which condition the full controllability solution is stable.
The Jacobian matrix, in this case simplify significantly and is given by 
\begin{equation}
	J=\left(\begin{array}{cccccccc}0&0&0&\frac{2P_{A}^{out}(2)}{\avg{k^{A}}_{out}}&0&0&0&0\\
		0&0&\frac{\Avg{k^{A}(k^{A}-1)}_{out}}{\avg{k^{A}}_{out}}&0& 0&0&0&0\\
		0& 0 & 0&0  & 0  &0&0&0\\
		\frac{\Avg{k^{A}(k^{A}-1)}_{in}}{\avg{k^{A}}_{in}}&0&0&0 &0& 0 &0&0\\
		0&0&0&0& 0&0&0&\frac{2P_{B}^{out}(2)}{\avg{k^{B}}_{out}}\\
		0&0&0&0 &0&0&\frac{\Avg{k^{B}(k^{B}-1)}_{out}}{\avg{k^{B}}_{out}}&0\\
		0 &0&0&0  &0& 0&0& 0\\
		0&0 &0&0  &\frac{\Avg{k^{B}(k^{B}-1)}_{in}}{\avg{k^{B}}_{in}}&0&0&0\\
	\end{array}\right)
\end{equation}
Four eigenvalues of $J$ are zero, the other four have degenerate modulus, therefore the stability conditions are
\bea\label{conditions}
2\frac{\Avg{k^{A}(k^{A}-1)}_{in}}{\avg{k^{A}}_{in}}\frac{P_{A}^{out}(2)}{\avg{k^{A}}_{out}}&<&1 \nonumber\\
2\frac{\Avg{k^{B}(k^{B}-1)}_{in}}{\avg{k^{B}}_{in}}\frac{P_{B}^{out}(2)}{\avg{k^{B}}_{out}}&<&1.
\eea
When $P_{A}^{in}(k)=P_{A}^{out}(k)=P_{B}^{in}(k)=P_{B}^{out}(k)=P(k)$ we have just one stability criterion  solution and it reads
\begin{equation}
	P(2)<\frac{\avg{k}^2}{2\Avg{k(k-1)}}
	\label{criterion_samepk}
\end{equation}
We observe here that on the single layers $\alpha=A,B$ the full controllability solution, is instead only stable  \cite{PRL} for 
\bea\label{conditions1layer}
2\frac{\Avg{k^{\alpha}(k^{\alpha}-1)}_{in}}{\avg{k^{\alpha}}_{in}}\frac{P_{\alpha}^{out}(2)}{\avg{k^{\alpha}}_{out}}&<&1 \nonumber\\
2\frac{\Avg{k^{\alpha}(k^{\alpha}-1)}_{out}}{\avg{k^{\alpha}}_{out}}\frac{P_{\alpha}^{in}(2)}{\avg{k^{\alpha}}_{in}}&<&1.
\eea
It follows that for duplex networks in which both layers have the same in-degree and out-degree distributions, i.e.  $P^{\alpha,in}(k)=P^{\alpha,out}(k)$ the stability of the full controllability solution on single layers is the same as the stability on the duplex network.
Nevertheless, for duplex networks formed by layers in which the in-degree distribution and the out-degree distribution are not the same there can be cases in which for the duplex network the fully controllable solution is stable while for the single layers it is not stable (See main text for discussion of this phenomenon and simulation results).

\subsubsection{Entropy}
In order to evaluate the number of maximum matchings, here we evaluate the entropy of the ground state solutions in the case of uncorrelated layers.
The entropy density is given by $s_0=S_0/N$ and can be computed by expanding the free energy at low temperatures $f(\beta\to \infty)=e_0-s_0/\beta+{\cal O}(1/\beta^2)$. This involves the study of the evanescent parts of the cavity field.
Therefore we assume that the field can be written as 
\bea
\begin{array}{ccc}
	h_{\alpha}=1+\frac{\ln \nu_{\alpha}}{\beta} &\mbox{for the peak around }& h=1\nonumber \\
	h_{\alpha}=-1+\frac{\ln \mu_{\alpha}}{\beta} &\mbox{for the peak around }& h=-1\nonumber \\
	h_{\alpha}=\frac{\ln \gamma_{\alpha}}{\beta} &\mbox{for the peak around }& h=0\nonumber \\
\end{array}
\eea
\bea
\begin{array}{ccc}
	\hat{h}_{\alpha}=1+\frac{\ln \hat{\nu}_{\alpha}}{\beta} &\mbox{for the peak around }& h=1\nonumber \\
	\hat{h}_{\alpha}=-1+\frac{\ln \hat{\mu}_{\alpha}}{\beta} &\mbox{for the peak around }& h=-1\nonumber \\
	\hat{h}_{\alpha}=\frac{\ln\hat{ \gamma}_{\alpha}}{\beta} &\mbox{for the peak around }& h=0\nonumber \\
\end{array}
\eea
From the BP equations, and the equations for $P(h_{\alpha})$ and $P(\hat{h}_{\alpha})$ we can  obtain the relation between the probability distributions ${\cal A}_{1}^{\alpha}(\nu_{\alpha}),{\cal A}_{2}^{\alpha}(\mu_{\alpha}),{\cal A}_{3}^{\alpha}(\gamma_{\alpha})$, and the distributions  $\hat{{\cal A}}_{1}^{\alpha}(\hat{\nu}_{\alpha}),\hat{{\cal A}}_{2}^{\alpha}(\mu_{\alpha}),\hat{{\cal A}}_{3}^{\alpha}(\hat{\gamma}_{\alpha})$, given by

\bea
{\cal A}_1^{\alpha}(\nu)&=&\sum_{k=0}^{\infty}\frac{ \left(\hat{w}_2^{\alpha}\right)^{k}}{w_1^{\alpha}}\frac{(k+1)}{\avg{k^{out}_{\alpha}}}P_{out}^{\alpha}(k+1)\int \left[\prod_{i=1}^{k}d{\hat{\mu}}_i^{\alpha}{\hat{\cal A}}_2^{\alpha}({\hat{\mu}}_i^{\alpha})\right]\delta\left(\nu-\frac{1}{1+\sum_{i=1}^{k}\hat{\mu}_i^{\alpha}}\right)\\
{\cal A}_2^{\alpha}(\mu)&=&
\sum_{k=1}^{\infty}\frac{1}{w_2^{\alpha}}\sum_{m=k}^{\infty}\frac{(m+1)}{\avg{k^{out}_{\alpha}}}P_{out}^{\alpha}(m+1)\binom{m}{k}(\hat{w}_1^{\alpha})^k(1-\hat{w}_1^{\alpha})^{m-k} \nonumber\\
&&\times\int\left[\prod_{i=1}^{k}d{\hat{\nu}}_i^{\alpha}{\hat{\cal A}}_1^{\alpha}({\hat{\nu}}_i^{\alpha})\right]\delta\left(\mu-\frac{1}{\sum_{i=1}^k \hat{\nu}_i^{\alpha}}\right) \\
{\cal A}_3^{\alpha}(\gamma)&=&\sum_{k=1}^{\infty}\sum_{m=k}^{\infty}\frac{1}{w_3^{\alpha}}\frac{(m+1)}{\avg{k^{out}_{\alpha}}}P_{out}^{\alpha}(m+1)\binom{m}{k}({\hat{w}}_3^{\alpha})^k(\hat{w}_2^{\alpha})^{m-k}\nonumber\\
&&\times\int\left[\prod_{i=1}^{k}d\hat{\gamma}_i^{\alpha}{\hat{\cal A}}_3^{\alpha}(\hat{\gamma}_i^{\alpha})\right]\delta\left(\gamma-\frac{1}{\sum_{i=1}^k \hat{\gamma}_i^{\alpha}}\right)
\eea

\bea
\hat{{\cal A}}_1^{A}(\hat{\nu})&=& \sum_{k^A=0}^{\infty}\frac{(k^A+1)}{\avg{k^{in}_{A}}}P_{in}^{A}(k^A+1)(w_2^A)^{k^A}\sum_{k^B=1}^{\infty}\sum_{m^B=k^B}^{\infty}P_{in}^{B}(m^B)\binom{m^B}{k^B}(w_1^B)^{k^B}(1-w_1^B)^{m^B-k^B}\nonumber \\
&&\int \left[\prod_{i=1}^{k^A}d{\mu}_i^A{{\cal A}}_2^{A}({\mu}_i^A)\prod_{i=1}^{k^B}d{\nu}_i^B{{\cal A}}_1^{B}({\nu}_i^B)\right]\times\delta\left(\hat{\nu}^A-\frac{1}{\frac{1}{\sum_{i=1}^{k^B}\nu_i^B}+\sum_{i=1}^{k^A}{\mu}_i^A}\right)
\eea
\bea
\hat{{\cal A}}_2^{A}(\hat{\mu})&=&\frac{1}{\hat{w}_2^{A}}G_1^{A,in}(1-w_1^A)\sum_{k^B}P_{in}^{B}(k^B)(w_2^B)^{k^B}\int\left[\prod_{i=1}^{k^B}d{\mu}_i^B{{\cal A}}_2^{B}({\mu}_i^B)\right]\delta\left(\hat{\mu}^A-\sum_{i=1}^{k^B}\mu^B_i\right)\nonumber \\
&+&\frac{1}{\hat{w}_2^{A}}(1-G_0^{B,in}(w_2^B))\sum_{k^A=1}^{\infty}\sum_{m^A=k^A}^{\infty}\frac{(m+1)}{\avg{k^{in}_{A}}}P_{in}^{A}(m^A+1)\binom{m^A}{k^A}(w_1^A)^k(1-w_1^A)^{m^A-k^A}\nonumber \\
&\times &\int\left[\prod_{i=1}^{k^A}d{\nu}_i^A{{\cal A}}_1({\nu}_i^A)\right]\delta\left(\hat{\mu}^A-\frac{1}{\sum_{i=1}^{k^A} \nu_i^A}\right)\nonumber \\
&+&\frac{1}{\hat{w}_2^{A}}\sum_{k^A=1}^{\infty}\sum_{m^A=k^A}^{\infty}\frac{(m+1)}{\avg{k^{in}_{A}}}P_{in}^{A}(m^A+1)\binom{m^A}{k^A}(w_1^A)^k(1-w_1^A)^{m^A-k^A}\sum_{k^B}P_{in}^{B}(k^B)(w_2^B)^{k^B}\nonumber \\
&\times &\int\left[\prod_{i=1}^{k^A}d{\nu}_i^A{{\cal A}}_1^{A}({\nu}_i^A)\prod_{i=1}^{k^B}d{\mu}_i^B{{\cal A}}_2^{B}({\mu}_i^B)\right]\delta\left(\hat{\mu}^A-\frac{1}{\frac{1}{\sum_{i=1}^{k^B}\mu_i^B}+\sum_{i=1}^{k^A}\nu_i^A}\right)
\eea
\bea
\hat{{\cal A}}_3^{A}(\hat{\gamma})&=&\frac{1}{\hat{w}_3^{A}}G_1^{A,in}(w_2^A)\sum_{k^B=1}^{\infty}\sum_{m^B=k^B}^{\infty}P_{in}^{B}(k^B)\binom{m^B}{k^B}(w_3^{B})^k(w_2^{B})^{m^B-k^B}\nonumber\\
&&\times\int\left[\prod_{i=1}^{k^B}d{\gamma}_i^B{{\cal A}}_3^{B}({\gamma}_i^B)\right]\delta\left(\gamma^A-\sum_{i=1}^{k_B} {\gamma}_i^B\right)\nonumber \\
&+&\frac{1}{\hat{w}_3^{A}}\sum_{k^A=1}^{\infty}\sum_{m^A=k^A}^{\infty}\frac{(m^A+1)}{\avg{k^{in}_{A}}}P_{in}^{A}(m^A+1)\binom{m^A}{k^A}(w_3^{A})^k(w_2^{A})^{m^A-k^A}\nonumber \\
&\times & \sum_{k^B=1}^{\infty}\sum_{m^B=k^B}^{\infty}P_{in}^{B}(k^B)\binom{m^B}{k^B}(w_3^{B})^k(w_2^{B})^{m^B-k^B}
\int\left[\prod_{i=1}^{k^A}d\gamma_i^{A}{\cal A}_3^{A}(\hat{\gamma}_i^{A})\prod_{i=1}^{k^B}d{\gamma}_i^B{{\cal A}}_3^{B}({\gamma}_i^B)\right]\nonumber \\
&\times &\delta\left(\gamma^A-\frac{1}{\frac{1}{\sum_{i=1}^{k_B} {\gamma}_i^B}+\sum_{i=1}^{k^A}\gamma_i^A}\right)\nonumber \\
&+&\frac{1}{\hat{w}_3^{A}}(1-G_0^{B,in}(1-w_1^B))\sum_{k^A=1}^{\infty}\sum_{m^A=k^A}^{\infty}\frac{(m^A+1)}{\avg{k^{in}_{A}}}P_{in}^{A}(m^A+1)\binom{m^A}{k^A}(w_3^{A})^k(w_2^{A})^{m^A-k^A}\nonumber \\
&\times & \int\left[\prod_{i=1}^{k^A}d\gamma_i^{A}{\cal A}_3^{A}(\hat{\gamma}_i^{A})\right]\delta\left(\gamma^A-\frac{1}{\sum_{i=1}^{k^A}\gamma_i^A}\right)
\eea

The free energy density $f(\beta)=F(\beta)/N=e_0-\frac{s_0}{\beta}+{\cal O}(1/\beta^2)$ with 
\begin{equation}
	s_0=s_{0,a,A}+s_{0,a,B}+s_{0,b}+s_{0,c,A}+s_{0,c,B}
	\label{entropyeq}
\end{equation}
where $s_{0,\ell}$ are given by 
\bea
s_{0,a,\alpha}&=&\sum_{k=1}^{\infty}\sum_{m=k}^{\infty}\binom{m}{k}(\hat{w}_1^{\alpha})^k (1-(\hat{w}_1^{\alpha}))^{m-k}P^{\alpha}_{out}(m)\overline{\ln\sum_{i=1}^k\hat{\nu}_i}\nonumber \\
&+&\sum_k P^{\alpha}_{out}(k) (\hat{w}_2^{\alpha})^k \overline{\ln \left(1+\sum_{i=1}^k \hat{\mu}_i\right)}\nonumber \\
&+& \sum_{k=1}^{\infty}\sum_{m=k}^{\infty}\binom{m}{k}(\hat{w}_3^{\alpha})^k(\hat{w}_2^{\alpha})^{m-k}P^{\alpha}_{out}(m)\overline{\ln\left(\sum_{i=1}^k\hat{\gamma}_i\right)}
\eea

\bea
s_{0,b}&=&(1-G_0^{B,in}(w_2^B))\sum_{k^A=1}^{\infty}\sum_{m^A=k^A}^{\infty}\binom{m^A}{k^A}(w_1^{A})^{k^A} (1-(w_1^A))^{m^A-k^A}P^{A}_{in}(m^A)\overline{\ln\sum_{i=1}^{k^A}\nu^A_i}\nonumber\\
&+&(1-G_0^{A,in}(w_2^A))\sum_{k^B=1}^{\infty}\sum_{m^B=k^B}^{\infty}\binom{m^B}{k^B}(w_1^{B})^{k^B} (1-(w_1^B))^{m^B-k^B}P^{B}_{in}(m^B)\overline{\ln\sum_{i=1}^{k^B}\nu^B_i}\nonumber\\
&+&(1-G_0^{B,in}(1-w_1^B))\sum_{k^A=1}^{\infty}\sum_{m^A=k^A}^{\infty}\binom{m^A}{k^A}(w_3^{A})^{k^A} (w_2^A)^{m^A-k^A}P^{A}_{in}(m^A)\overline{\ln\sum_{i=1}^{k^A}\gamma^A_i}\nonumber\\
&+&(1-G_0^{A,in}(1-w_1^A))\sum_{k^B=1}^{\infty}\sum_{m^B=k^B}^{\infty}\binom{m^B}{k^B}(w_3^{B})^{k^B} (w_2^B)^{m^B-k^B}P^{B}_{in}(m^B)\overline{\ln\sum_{i=1}^{k^B}\gamma^B_i}\nonumber\\
&+& \left[\sum_{k^B}P^{B}_{in}(k^B)(w_2^B)^{k^B}\sum_{k^A=1}^{\infty}\sum_{m^A=k^A}^{\infty}\binom{m^A}{k^A}(w_1^{A})^{k^A} (1-(w_1^A))^{m^A-k^A}P^{A}_{in}(m^A)\right.\nonumber\\
&&\left. \times \overline{\ln{ \left(1+\sum_{i=1}^{k^A}\nu^A_i\sum_{i=1}^{k^B}\mu^B_i\right)}}\right]\nonumber\\
&+&\left[\sum_{k^A}P^{A}_{in}(k^A)(w_2^A)^{k^A}\sum_{k^B=1}^{\infty}\sum_{m^B=k^B}^{\infty}\binom{m^B}{k^B}(w_1^{B})^{m^B} (1-(w_1^B))^{m^B-k^B}P^{B}_{in}(m^B)\right.\nonumber\\
&&\left. \times\overline{\ln{ \left(1+\sum_{i=1}^{k^B}\nu^B_i\sum_{i=1}^{k^A}\mu^A_i\right)}}\right]\nonumber\\
&+&\left[\sum_{k^A=1}^{\infty}\sum_{m^A=k^A}^{\infty}\binom{m^A}{k^A}(w_3^{A})^{k^A} (w_2^A)^{m^A-k^A}P^{A}_{in}(m^A)\right.\nonumber\\
&&\left. \times\sum_{k^B=1}^{\infty}\sum_{m^B=k^B}^{\infty}\binom{m^B}{k^B}(w_3^{B})^{k^B} (w_2^B)^{m^B-k^B}P^{B}_{in}(m^B)\overline{\ln{\left(1+\sum_{i=1}^{k^A}\gamma^A_i\sum_{i=1}^{k^B}\gamma^B_i\right)}}\right]
\eea

%%%%%%%%%
\bea
s_{0,c,\alpha}&=&-\avg{k^{\alpha}}_{in}\left\{\hat{w}_1^{\alpha}(w_1^{\alpha}+w_3^{\alpha})\overline{\ln \hat{\nu}^{\alpha}}+{w}_1^{\alpha}(\hat{w}_1^{\alpha}+\hat{w}_3^{\alpha})\overline{\ln {\nu}^{\alpha}}\right.\nonumber \\
&+&\hat{w}_1^{\alpha}w_2^{\alpha}\overline {\ln(1+\hat{\nu}\mu)}+{w}_1^{\alpha}\hat{w}_2^{\alpha}\overline{\ln (1+{\nu}\hat{\mu})}\nonumber \\
&+&\left.\hat{w}_1^{\alpha}w_3^{\alpha}\overline{\ln \gamma}+{w}_1^{\alpha}\hat{w}_3^{\alpha}\overline{\ln \hat{\gamma}}+w_3^{\alpha}\hat{w}_3^{\alpha}\overline{\ln (1+\hat{\gamma}\gamma)}\right\}.
\eea

\subsubsection{Phase transition in the controllability of Poisson duplex networks.}
\label{phasetpoiss}
\begin{figure}[!t]
	\begin{center}
		{\includegraphics[width=0.99\columnwidth]{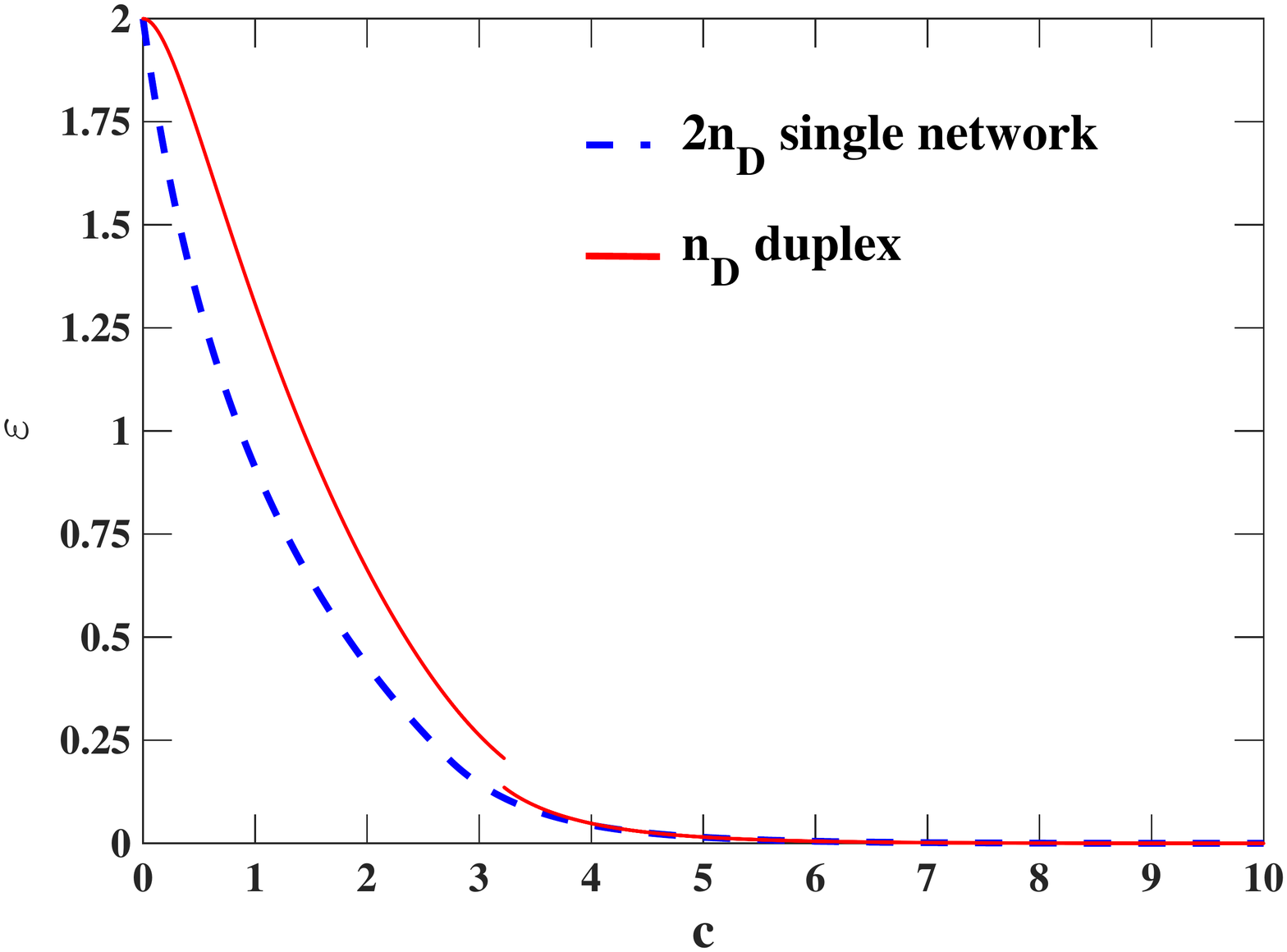}}
	\end{center}
	\caption{Density of driver nodes $\varepsilon=n_D$  for a duplex network composed by two Poisson networks with $\Avg{k^{A,in}}=\Avg{k^{A,out}}=\Avg{k^{B,in}}=\Avg{k^{B,out}}=c$ is indicated with a solid red line and it clearly shows a phase transition for $c=3.22233\ldots$.  In the dashed blue line we display the double of the number of driver nodes $\varepsilon=2n_D$  for a single Poisson network with the same average degree $c$, indicating the fraction of  driver nodes necessary to control separately the two layers.
	} 
	\label{comparisonduplex1netpoisson}
\end{figure}
Here we consider the case  of two Poisson networks with the same in/out average degree. 
In other words, we consider the situation in which
$\Avg{k^{A,in}}=\Avg{k^{A,out}}=\Avg{k^{B,in}}=\Avg{k^{B,out}}=c$.
The fraction $n_D$ of  nodes that are driver nodes of this duplex, is always larger than the double of the fraction of driver nodes in each of the layers taken in isolation (see   Fig. \ref{comparisonduplex1netpoisson}). Moreover we observe that  there is a phase transition in the controllability of these duplex networks, indicated by a discontinuity of $n_D$ for  $c=c^{\star}=3.22233\ldots$ (see   Fig. \ref{comparisonduplex1netpoisson}).
In order to derive these results,  we  assumed $w_i^A=w_i^B$ for $i=1,2,3$ and $\hat{w}_i^A=\hat{w}_i^B$ for $i=1,2,3$. Therefore  the zero-temperature BP equations at the ensemble level 
$(\ref{equncorr})$ read,
\bea
w_1&=&e^{-c(1-\hat{w}_2)},\nonumber \\
w_2&=&\left[1-e^{-c\hat{w}_1}\right],\nonumber \\
w_3&=&1-w_1-w_2,\nonumber \\
\hat{w}_3&=&1-\hat{w}_1-\hat{w}_2,\nonumber\\
\hat{w}_1&=&e^{-c(1-w_2)}\left[1-e^{-c{w}_1}\right],\nonumber \\ 
\hat{w}_2&=&\left[1-e^{-c w_1}+e^{-cw_1}e^{-c(1-w_2)}\right].\nonumber \\
\eea

The energy $E$ is given in this case by 
\bea
E&=&2\left[e^{-c(1-\hat{w}_2)}-1+e^{-c\hat{w}_1}\right]-2[1-e^{-cw_1}][1-e^{-c(1-w_2)}]+2c\left[\hat{w}_{1}(1-w_2)+w_{1}(1-\hat{w}_2)\right].
\eea

\begin{figure*}
	\center
	\includegraphics[width=\textwidth]{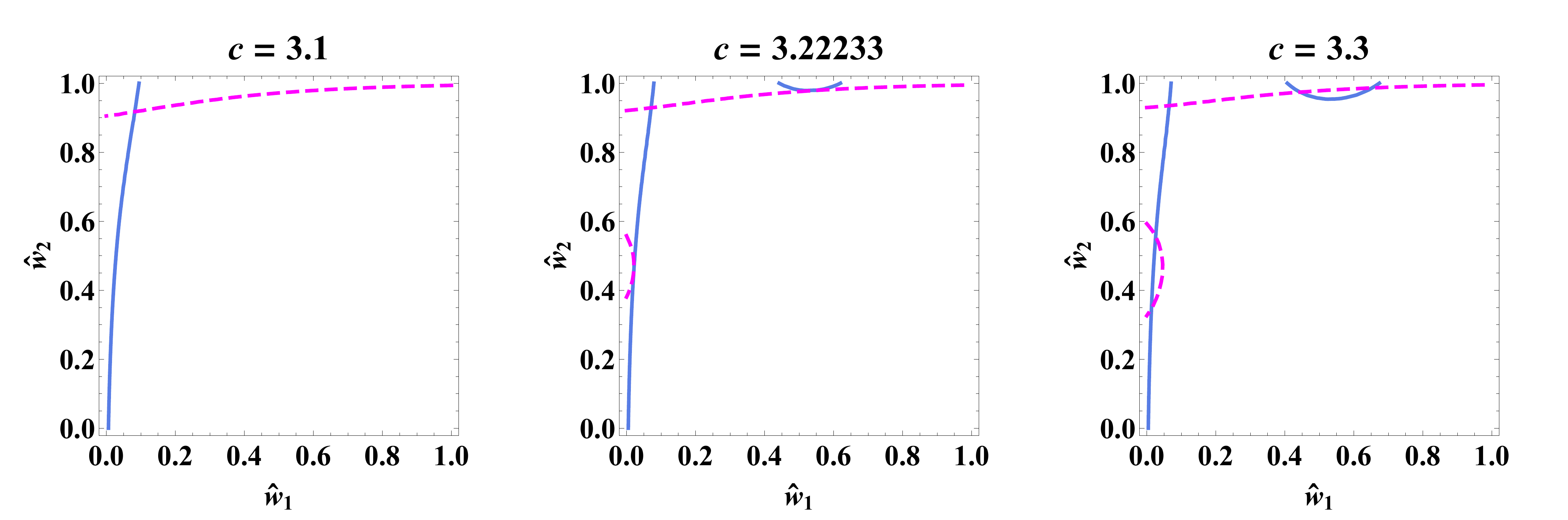}
	\caption{ Plots of the functions $\hat{w}_1=h_1(\hat{w}_1,\hat{w}_2)$ and $\hat{w}_2=h_2(\hat{w}_1,\hat{w}_2)$ given by Eqs. $( \ref{2w1hatsameDNsupp})-(\ref{2w2hatsameDNsupp})$.  The   solution of the system of these two equations, corresponds to a crossing of the two curves. We show the emergence of two new solutions of this system  of equations for $c>c^{\star}=3.22233\ldots$. The critical point $c^{\star}$ characterize an hybrid phase transition in the controllability of the duplex network.} 
	\label{fig3supp}
\end{figure*}

%%%%%%%%%%%%%%%%%%%%%%%%%%%%%%%%%%%%%%%%%%%%%%%%%%%%%%%%%%%
\begin{figure}
	\begin{center}
		{\includegraphics[width=0.99\columnwidth]{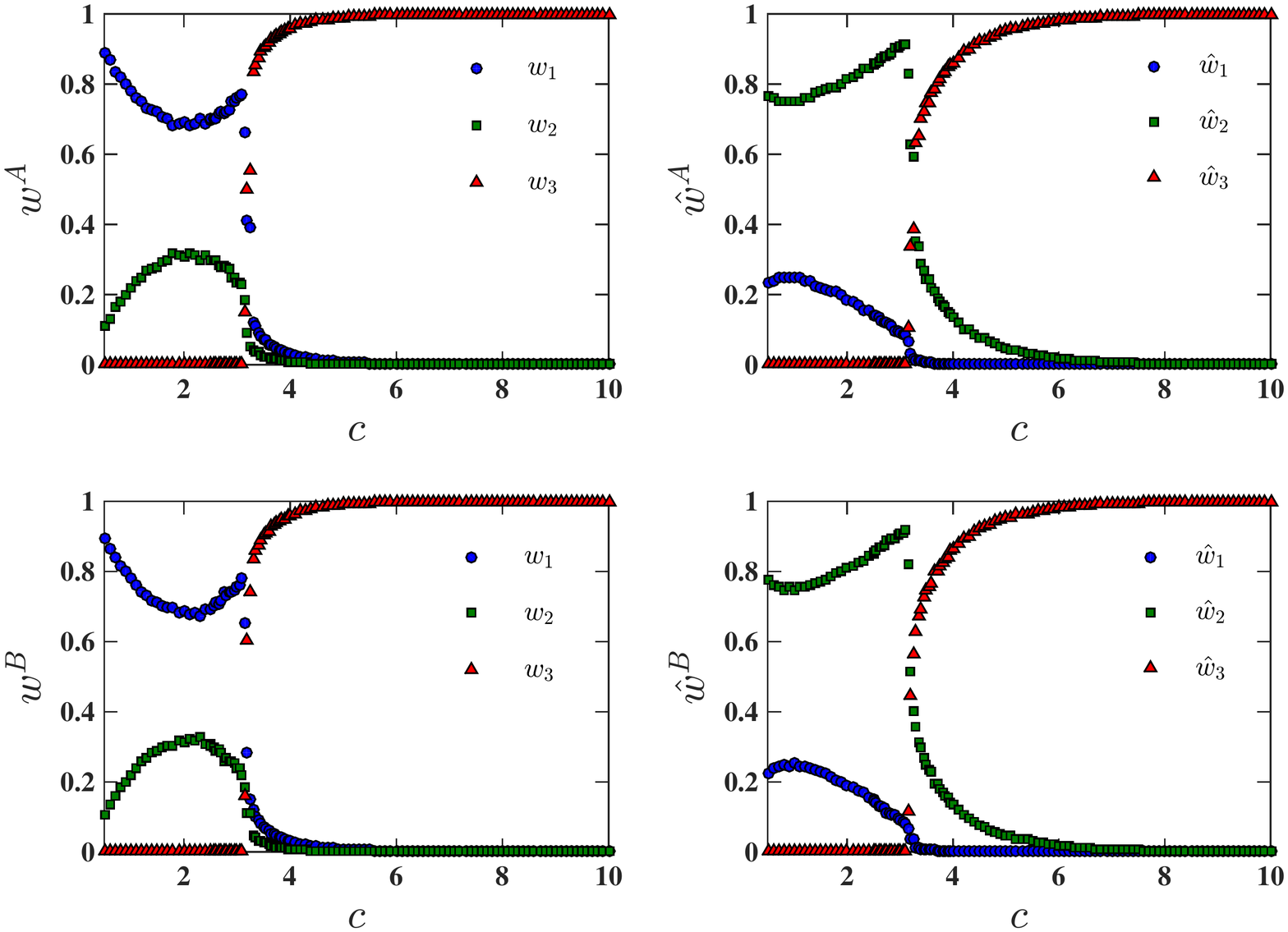}}
	\end{center}
	\caption{Values of the probabilities $\{w_i\}_{i=1,2,3}$ and $\hat{w}_{i=1,2,3}$  plotted as a function of the average degree $c$, for a  duplex network formed by two Poisson layers with $\Avg{k^{A,in}}=\Avg{k^{A,out}}=\Avg{k^{B,in}}=\Avg{k^{B,out}}=c$.   These probabilities are calculated directly from BP results obtained over 5  single realizations these multiplex networks with average degree $c$ and  $N=10^4$.
		%{\bf change figure to be readable in black and white, use different symbols}
	} 
	\label{wpoissonpaper}
\end{figure}

We notice that the equations for $\hat{w}_1$ and $\hat{w}_2$ can be rewritten to form a closed subsystem of equations, 
\begin{subequations}
	\bea
	\hat{w}_1&=&h_1(\hat{w}_1,\hat{w}_2)=e^{-ce^{-c\hat{w}_1}}\left[1-e^{-ce^{-c(1-\hat{w}_2)}}\right] 
	\label{2w1hatsameDNsupp}\\ 
	\hat{w}_2&=&h_2(\hat{w}_1,\hat{w}_2)=\left[1-e^{-c e^{-c(1-\hat{w}_2)}}+e^{-ce^{-c(1-\hat{w}_2)}}e^{-ce^{-c
			\hat{w}_1}}\right] \label{2w2hatsameDNsupp}
	\eea\label{hatsameDNsupp}
\end{subequations}
from the solution of which the remaining quantities can be determined.
%
%\bea
%\hat{w}_1&=&e^{-ce^{-c\hat{w}_1}}\left[1-e^{-ce^{-c(1-\hat{w}_2)}}\right], 
%\label{w1hatsameDN}\\ 
%\hat{w}_2&=&\left[1-e^{-c e^{-c(1-\hat{w}_2)}}+e^{-ce^{-c(1-\hat{w}_2)}}e^{-ce^{-c
%\hat{w}_1}}\right], \label{w2hatsameDN}\\
%w_1&=&e^{-c(1-\hat{w}_2)},\nonumber \\
%w_2&=&\left[1-e^{-c\hat{w}_1}\right],\nonumber \\
%w_3&=&1-w_1-w_2,\nonumber \\
%\hat{w}_3&=&1-\hat{w}_1-\hat{w}_2.\nonumber
%\eea

The value $c^{\star}$ of the average degree $c$ at which the discontinuity in the number of driver nodes $n_D$ observed in Fig.\ref{comparisonduplex1netpoisson} occurs can be found by imposing that the two curves $\hat{w}_1= h_1(\hat{w}_1,\hat{w}_2)$ and $\hat{w}_2= h_2(\hat{w}_1,\hat{w}_2)$ of the plane $w_1,w_2$ for $c=c^{\star}$ are tangent to each other at their interception. These functions are plotted in Figure $\ref{fig3supp}$ where it is possible to observe that for $c>c^{\star}$ the curves cross in three  points while for $c<c^{\star}$ they  cross in one point, and at $c=c^{\star}$ they are tangent to each other.
The critical  point $c^{\star}$ is found by imposing that the Eqs.~\eqref{hatsameDNsupp} are satisfied together with the condition
\bea
\left|J\right|=0,
\label{Jsupp}
\eea
with $J$ indicating the Jacobian of the system of equations $\hat{w}_1=h_1(\hat{w}_1,\hat{w}_2)$ and $\hat{w}_2=h_2(\hat{w}_1,\hat{w}_2)$ given by 
\bea
J=\left(\begin{array}{cc}1-\frac{\partial h_1(\hat{w}_1,\hat{w}_2)}{\partial \hat{w}_1}&-\frac{\partial h_1(\hat{w}_1,\hat{w}_2)}{\partial \hat{w}_2} \nonumber \\
	-\frac{\partial h_2(\hat{w}_1,\hat{w}_2)}{\partial \hat{w}_1}&1-\frac{\partial h_2(\hat{w}_1,\hat{w}_2)}{\partial \hat{w}_2} \end{array}\right).
\eea
Imposing that Eqs.~\eqref{hatsameDNsupp} and condition \eqref{Jsupp} are simultaneously satisfied, the solution $c^{\star}=3.222326106\ldots$ is found.
For $c<c^{\star}$ we observe that $w_3=\hat{w}_3=0$. At $c^{\star}$ we observe a discontinuity in both $w_3$ and $\hat{w}_3$, but for $c>c^{\star}$ the functions $h_1(\hat{w}_1,\hat{w}_2)$ and $h_2(\hat{w}_1,\hat{w}_2)$ are analytic, and analyzing Eqs. $(\ref{2w1hatsameDNsupp})-(\ref{2w2hatsameDNsupp})$ we obtain the behavior of the order parameters $w_3$ and $\hat{w}_3$ for $c>c^{\star}$
\bea
w_3-w_3^{\star}&\propto &(c-c^{\star})^{1/2}\nonumber \\
\hat{w}_3-\hat{w}_3^{\star}&\propto &(c-c^{\star})^{1/2},
\eea
showing that the transition is hybrid.\\

%\end{document}
% Whenever $P_{A}^{in}(k)=P_{A}^{out}(k)=P_{B}^{in}(k)=P_{B}^{out}(k)=P(k)$ and the minimum degree is 2,  the solution $w^{\alpha}_1=\hat{w}^{\alpha}_1=w^{\alpha}_2=\hat{w}^{\alpha}_2=0$ and $w_3^{\alpha}=\hat{w}^{\alpha}_3=1$ is associated with just one stability criterion, i.e. Eq. \ref{criterion_samepk}. 

\begin{figure}
	\begin{center}
		{\includegraphics[width=0.99\columnwidth]{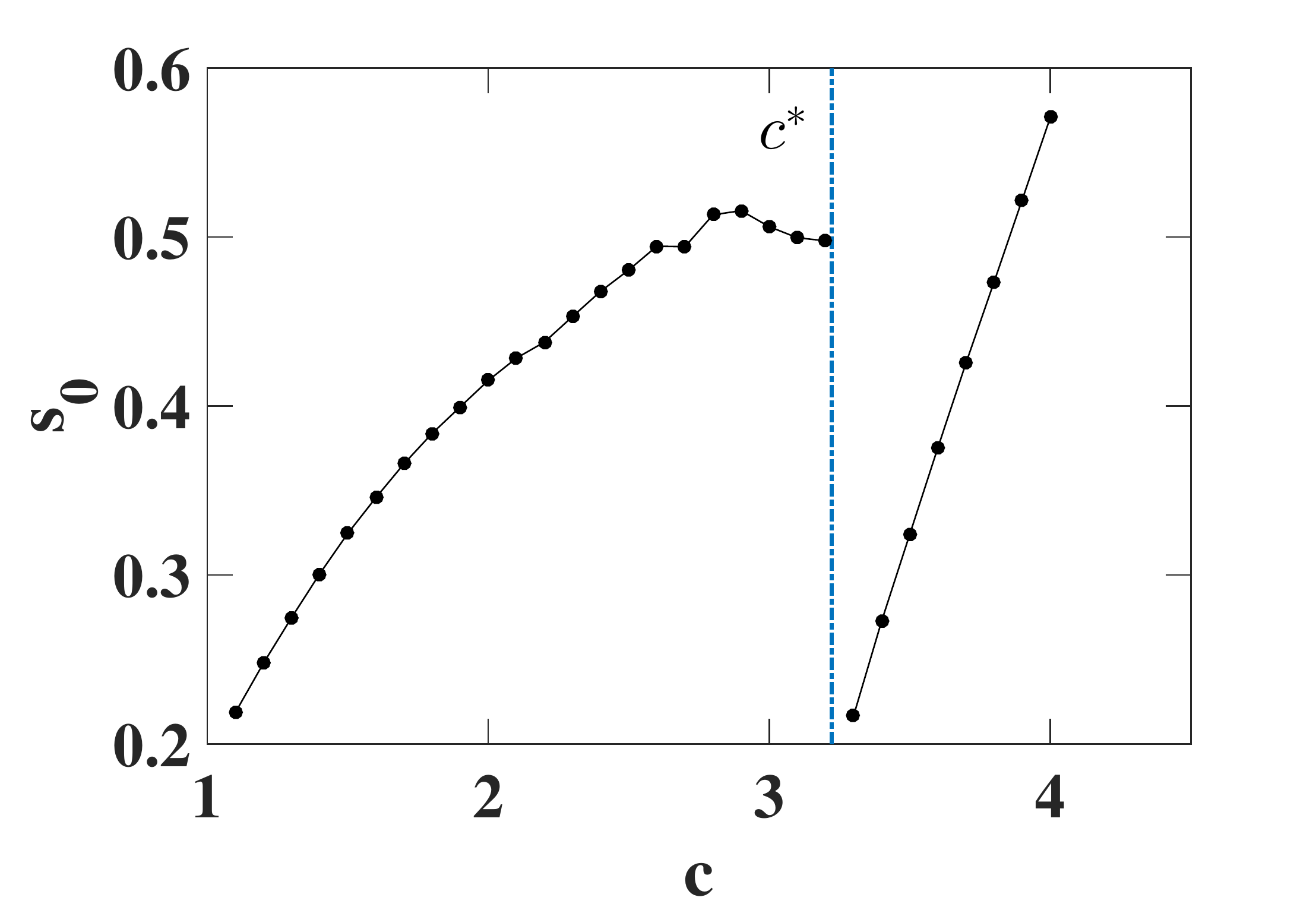}}
	\end{center}
	\caption{Entropy density $s$ for a duplex network composed by two Poisson networks with $\Avg{k^{A,in}}=\Avg{k^{A,out}}=\Avg{k^{B,in}}=\Avg{k^{B,out}}=c$. At $c^*=3.22233..$, average degree corresponding to the hybrid phase transition,  the entropy density $s$ displays a finite jump.  }
	\label{entropypic}
\end{figure}

We further characterize this phase transition evaluating  the number of maximum matchings, i.e. the entropy value of the ground state solutions in the case of two poisson uncorrelated layers. The entropy density $s$ follows from Eq. (\ref{entropyeq}) and it is plotted as a function of the average degree $c$ in Fig. \ref{entropypic}. The entropy density presents a small jump at $c^*=3.22233..$ marking a change in the properties of the solutions.
\\

\begin{figure}[!t]
	\begin{center}
		{\includegraphics[width=0.99\columnwidth]{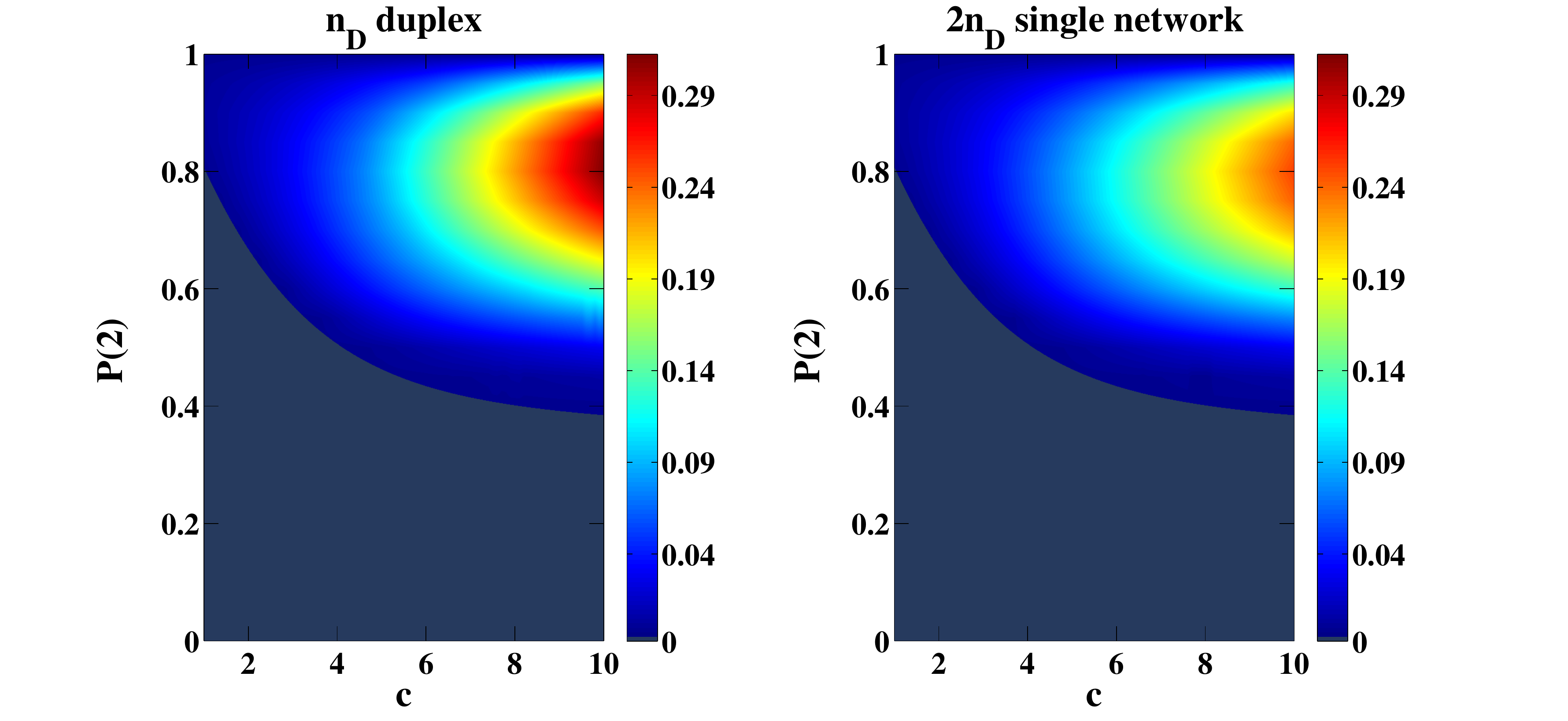}}
	\end{center}
	\caption{On the left: the density of driver nodes $n_D$  as a function of the parameters $c$ and $P(2)$ is plotted for duplex networks with $P_{A}^{in}(k)=P_{A}^{out}(k)=P_{B}^{in}(k)=P_{B}^{out}(k)=P(k)$ and $P(k)$ given by Eq. $(\ref{pkinoutpoisson})$. On the right:  the double of the  density of driver nodes $n_D$  for single layers with degree distribution  $P^{in}(k)=P^{out}(k)=P(k)$ and $P(k)$ given by $(\ref{pkinoutpoisson})$ is  plotted  as a function of $c$ and $P(2)$.} 
	\label{comparisonpoissonduplex1net_P2var}
\end{figure}

Here we want to modify the degree distribution of the duplex network characterized in this section, by changing the probability of  nodes of low degree (degree $0,1,2$) that have been shown to be essential to determine the controllability of single layers \cite{PRL}.
Therefore we consider  a duplex networks  with degree distributions  
$P_{A}^{in}(k)=P_{A}^{out}(k)=P_{B}^{in}(k)=P_{B}^{out}(k)=P(k)$ and with  minimum degree is 2.
In particular we consider $P(k)$ given by  
\bea
P(k)=\left\{\begin{array}{lcc}
	0 &\mbox{for}&{k<2}\\
	P(2)&\mbox{for}& k=2 \\
	\kappa\frac{1}{k!}c^k &\mbox{for} &k\in [3,\infty]\end{array}\right.
\label{pkinoutpoisson}
\eea
with $\kappa$ indicating a  normalization constant. In Fig.~\ref{comparisonpoissonduplex1net_P2var} (on the left) we show the phase diagram of this  duplex network described by the dependence of the fraction of driver nodes $n_D$ on  $c$ and $P(2)$. The dark grey area  defines the region where the zero-energy solution is stable, hence in which to control a duplex one needs only an infinitesimal fraction of driver nodes, i.e. $n_D=0$.  These results are compared with the situation in which the two layers are controlled separately shown in  Fig. \ref{comparisonpoissonduplex1net_P2var} (on the right). The fraction of driver replica nodes of the duplex network is always larger that the double of the fraction of driver nodes in any single layer taken in isolation. Moreover the region in which the fully controllable solution is stable is the same for the duplex network, and for the single networks in the layers of the duplex network taken in isolation. This result is  consistent with the  theoretical expectations obtained in Sec.~\ref{stabcondduplex}.  In fact the in- and out-degree distributions of the two layers are the same.

\subsubsection{Controllability of scale-free duplex networks}
\begin{figure}[!t]
	\begin{center}
		{\includegraphics[width=0.99\columnwidth]{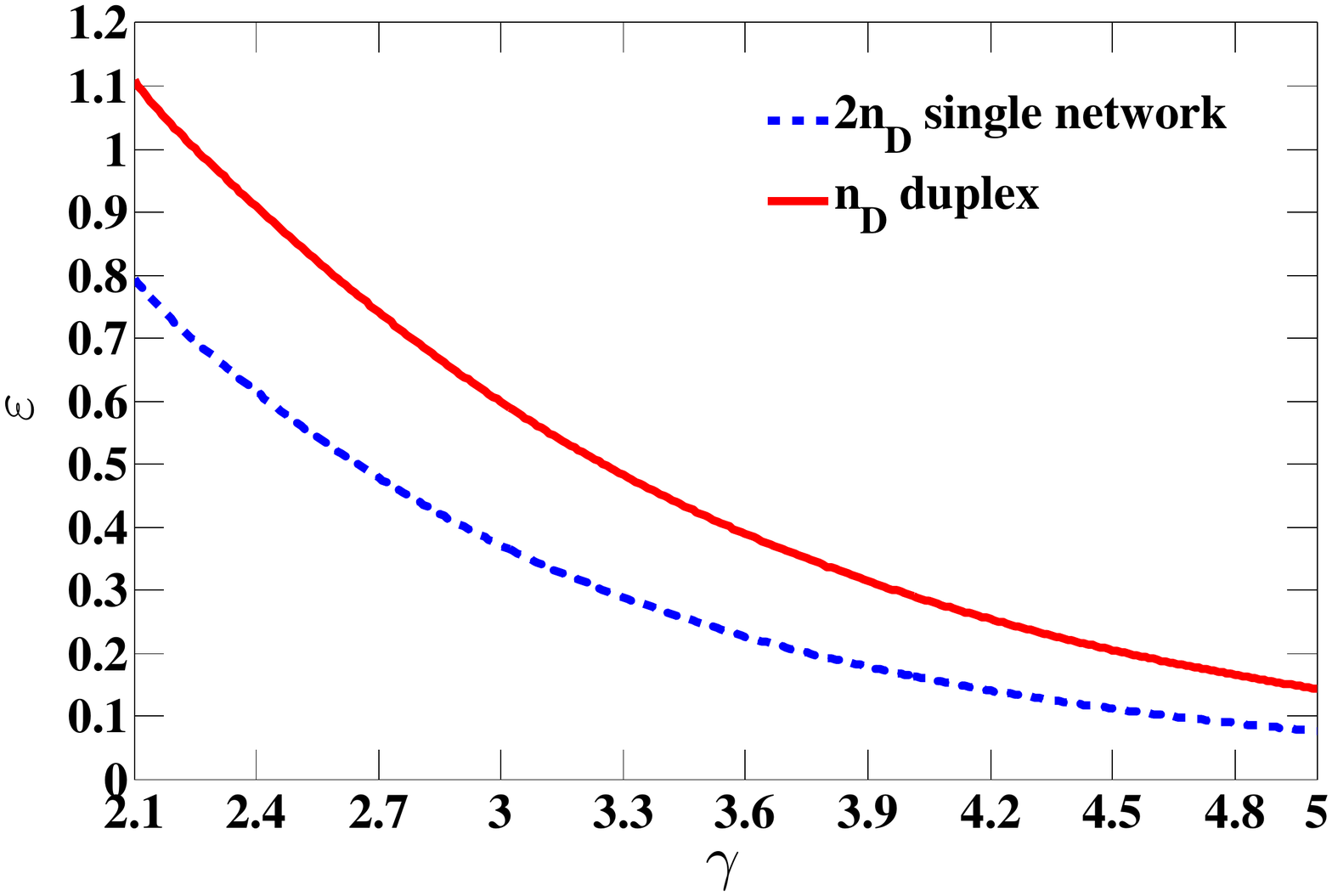}}
	\end{center}
	\caption{Density of driver nodes $\varepsilon=n_D$  for a duplex network composed by two power-law networks with $P(k^{A,in})=P(k^{A,out})=P(k^{B,in})=P(k^{B,out})=P(k)\propto k^{-\gamma}$ and minimal degree $m=1$ as a function of $\gamma$ (indicated with a solid red line). The minimum in/out degree $1$ and the maximum in/out degree is given by the structural cutoff with $N=10^4$. In the dashed blue line we display the double of the number of driver nodes $\varepsilon=2n_D$  for a single power-law network with the same in/out degree distributions $P(k)$, indicating the fraction of  driver nodes necessary to control separately the two layers.
	} 
	\label{plotpowerlawcomparisonduplex1net}
\end{figure}

Following Sec. \ref{phasetpoiss} we consider now the case of two uncorrelated layers composed by two power-law networks with $P(k^{A,in})=P(k^{A,out})=P(k^{B,in})=P(k^{B,out})=P(k)\propto k^{-\gamma}$ and minimal degree $m=1$.
Similarly to the poisson case, the fraction $n_D$ of driver nodes of this duplex, is always larger than the double of the fraction of driver nodes in each of the layers taken in isolation (see Fig. \ref{plotpowerlawcomparisonduplex1net}).

\begin{figure}[!t]
	\begin{center}
		{\includegraphics[width=0.99\columnwidth]{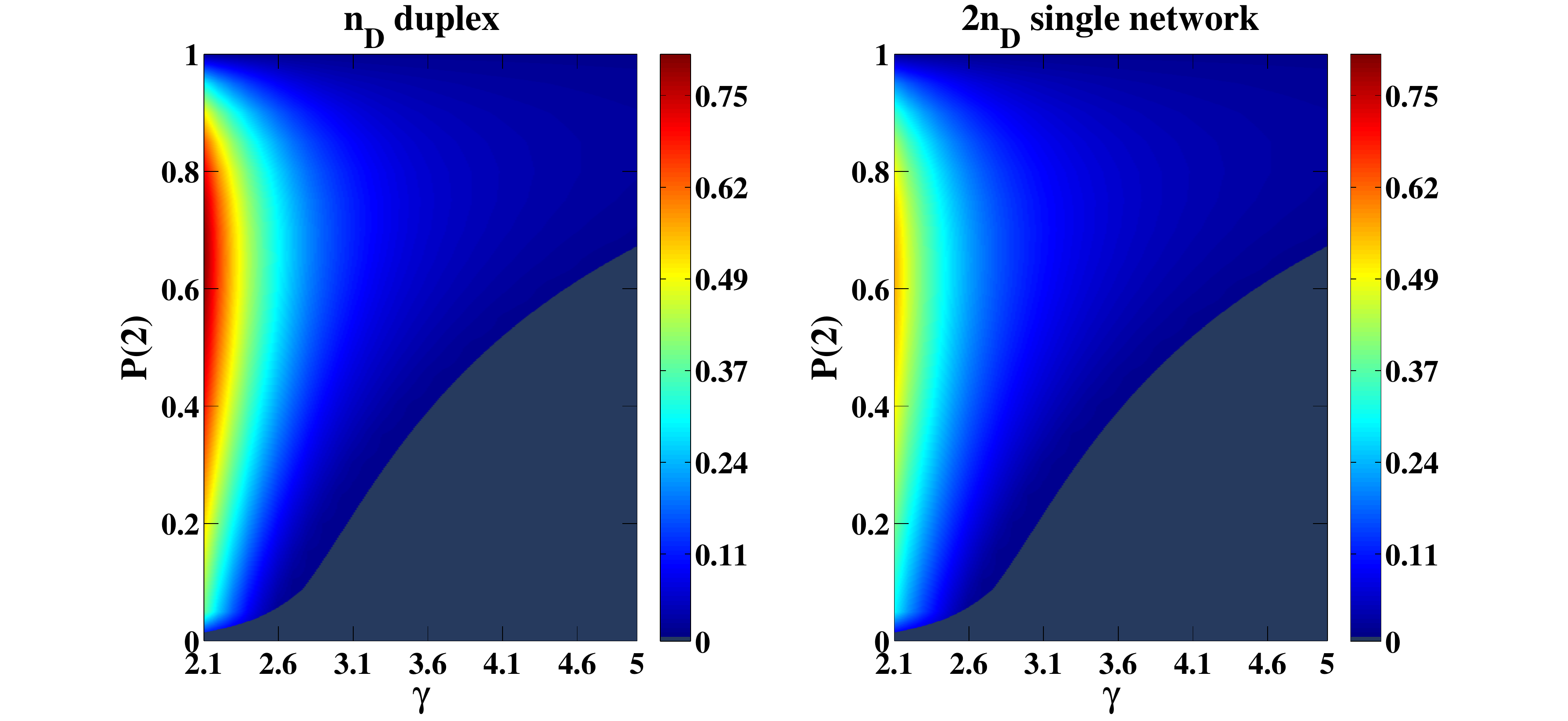}}
	\end{center}
	\caption{On the left: the density of driver nodes $n_D$  as a function of the parameters $\gamma$ and $P(2)$ for duplex networks of $N=10^6$ nodes with degree distributions $P_{A}^{in}(k)=P_{A}^{out}(k)=P_{B}^{in}(k)=P_{B}^{out}(k)=P(k)$ and $P(k)$ given by Eq. $(\ref{pkinout})$. On the right: double of  the density of driver nodes $n_D$  as a function of the parameters $\gamma$ and $P(2)$ for single networks  of $N=10^6$ nodes with degree distributions $P^{in}(k)=P^{out}(k)=P(k)$  and $P(k)$ given by Eq. $(\ref{pkinout})$.} 
	\label{comparisonduplex1netp2gamma}
\end{figure}

\begin{figure}[!t]
	\begin{center}
		{\includegraphics[width=0.99\columnwidth]{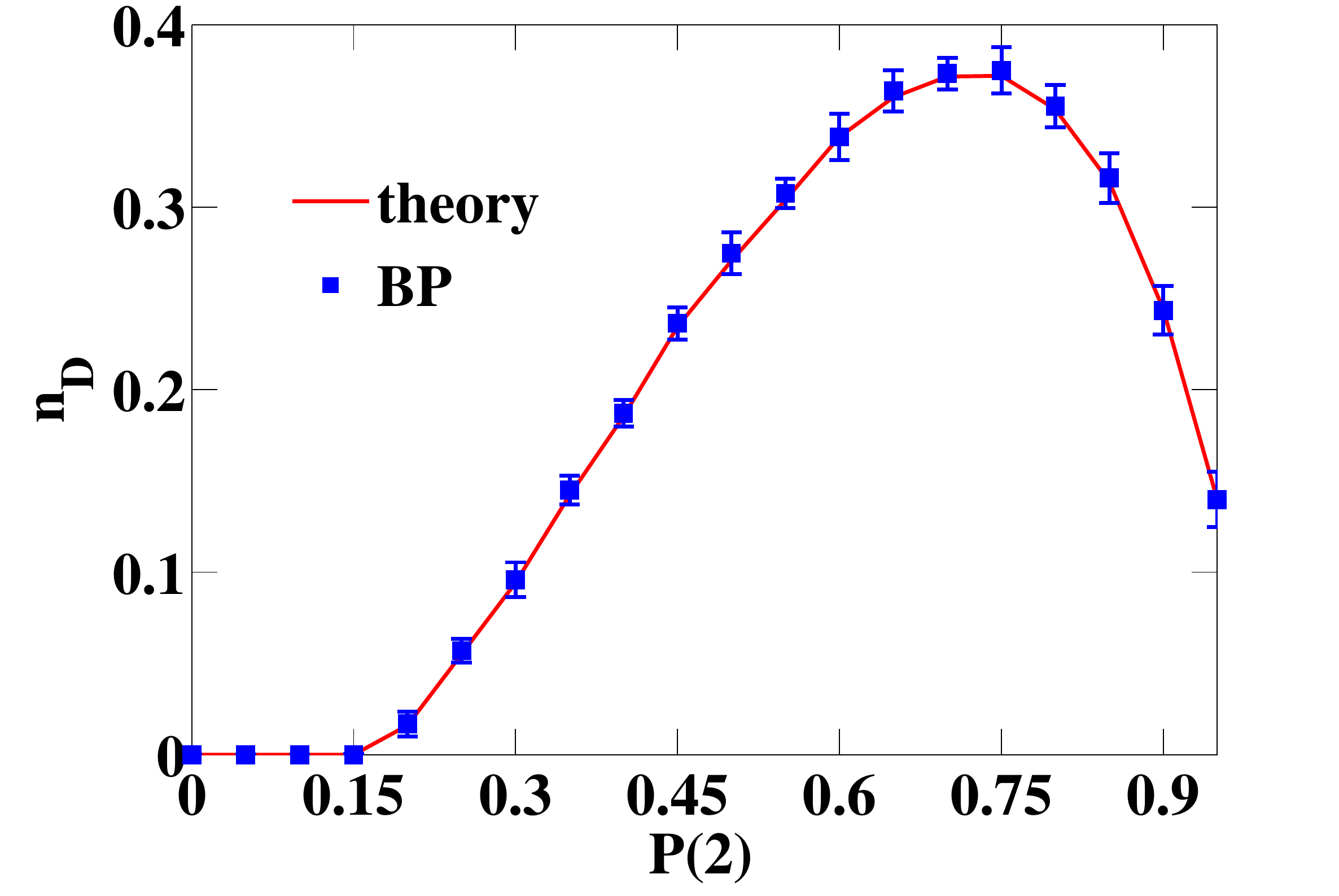}}
	\end{center}\caption{Density of driver nodes $n_D$ as a function of $P(2)$ for a duplex network with $P_{A}^{in}(k)=P_{A}^{out}(k)=P_{B}^{in}(k)=P_{B}^{out}(k)=P(k)$, $P(k)$ given by  Eq. $(\ref{pkinout})$  and $\gamma=2.3$. The fraction of driver nodes computed with the zero-temperature BP (Max-Sum) algorithm on a duplex network of $N=10^4$ nodes (averaged over $25$ network realizations) is compared with the theoretical expectation for the density $n_D$ in an ensemble of random duplex networks with the given degree distributions.
} 
\label{P2varduplex}
\end{figure}

Moreover, low-degree nodes significantly affect  the controllability of duplex networks formed by scale-free networks.
We consider a duplex network with  $P_{A}^{in}(k)=P_{A}^{out}(k)=P_{B}^{in}(k)=P_{B}^{out}(k)=P(k)$ and $P(k)$ given by 
\bea
P(k)=\left\{\begin{array}{lcc}
	0 &\mbox{if}&{k=1}\\
	P(2)&\mbox{if}& k=2 \\
	\kappa k^{-\gamma} &\mbox{if} &k\in [3,M]\end{array}\right.
\label{pkinout}
\eea
with $\kappa$ indicating the normalization sum and $\gamma>2$.
We consider uncorrelated networks, therefore the cutoff $M$ on the degrees of the nodes will be given by 
\bea
M=\min(\sqrt{N},\left\{[1-P(1)-P(2)]N\right\}^{1/(\gamma-1)}).
\eea
In other words, the cutoff $M$ is given by the minimum between the structural cutoff of the network and the natural cutoff of the degree distribution.
In Fig. \ref{comparisonduplex1netp2gamma} (on the left) we present the phase diagram of a duplex network displaying the fraction of driver nodes $n_D$ as a function of the parameters $\gamma$ and $P(2)$. The dark grey area is associated with the stable zero-energy solution while outside this region, the minimum fraction of driver nodes necessary for a full duplex control follows the colorcode. We compare these results with the situation in which the two layers are controlled separately (on the right).
We observe that the number of driver replica nodes in the duplex is always greater than the total number of driver nodes of the single layer taken in isolation, provided that the duplex network is not fully controllable.
We note that for the degree distribution considered in this case, consistently with the theoretical results obtained in Sec. $\ref{stabcondduplex}$, we observe that the region for the stability of the full controllability solution for  the duplex network is the same of the region for the stability of the full controllability solution in the single layers. 
Finally , in Fig. \ref{P2varduplex} we compare our theoretical results for the ensemble of duplex networks with  degree distributions  $P_{A}^{in}(k)=P_{A}^{out}(k)=P_{B}^{in}(k)=P_{B}^{out}(k)=P(k)$ and  $P(k)$ given by  Eq. $(\ref{pkinout})$, with those obtained by the message-passing (BP) algorithm, finding a  good agreement (Eq. \ref{criterion_samepk} returns a limit value for $P(2)$ equal to $0.181947$).

\subsection {Effect of degree correlations on controllability of multiplex networks}

In order to analyze the effect of degree correlations \cite{PhysReports} on the controllability of multiplex  networks, we correlate the degree of the replica nodes in the two layers of a duplex network formed by layer $A$ and layer $B$.
In particular we consider two cases: a duplex network in which only the low in-degree nodes (nodes of in-degree $0,1,2$ ) are correlated and a duplex networks in which the in-degrees of the replica nodes are correlated independently on their value.
For each case, we define the joint in-degree distribution  $P^{in}(k^A,k^{B})$ between layers and the corresponding expression of the zero-temperature BP equations in the correlated ensemble of networks.

In the first case we consider a joint in-degree distribution $P^{in}(k^A,k^{B})$ given by 
\bea
P_{in}(k^A,k^B) =\left\{ \begin{array}{lll} p\delta_{k^B,k^A}P(k^A)+(1-p)P(k^A)P(k^B), & \mbox{for}& k^A\le 2  \nonumber \\ (1-p)P(k^A)P(k^B), & \mbox{for }& k^A > 2 \quad k^B \le 2\nonumber  \\ p\frac{P(k^B)}{C}P(k^A)+(1-p)P(k^A)P(k^B), & \mbox{for }& k^A > 2 \quad k^B > 2 \end{array}\right.,
\eea
where $C=1-\sum_{k\le 2} P(k)$ where $P(k)$ is a given normalized degree distribution. 
The distributions of the fields over the links of this  ensemble of networks are given by
\bea
{\cal P}_{\alpha}(h^{\alpha})&=&w_1^{\alpha}\delta(h^{\alpha}-1)+w_2^{\alpha}\delta(h^{\alpha}+1)+w_3^{\alpha}\delta(h^{\alpha}),\nonumber \\
\hat{{\cal P}}_{\alpha}(\hat{h}^{\alpha})&=&\hat{w}_1^{\alpha}\delta(\hat{h}^{\alpha}-1)+\hat{w}_2^{\alpha}\delta(\hat{h}^{\alpha}+1)+\hat{w}_3^{\alpha}\delta(\hat{h}^{\alpha}),
\eea
where $\alpha=A,B$  and where the probabilities $w_1^{\alpha},w_2^{\alpha},w_3^{\alpha}$ are normalized $w_1^{\alpha}+w_2^{\alpha}+w_3^{\alpha}=1$ as well as the probabilities $\hat{w}_1^{\alpha},\hat{w}_2^{\alpha},\hat{w}_3^{\alpha}$ that satisfy the equation $\hat{w}_1^{\alpha}+\hat{w}_2^{\alpha}+\hat{w}_3^{\alpha}=1$.
The zero-temperature BP (Max-Sum) equations \eqref{maxsum} averaged over this ensemble of networks can be expressed in terms of the probabilities 
$\{w_i^{\alpha}\}_{i=1,2,3}$ and $\{w_i^{\alpha}\}_{i=1,2,3}$ as
\begin{subequations}
	\bea
	\hat{w}_1&=&p\left[\frac{P(1)}{\avg{k}}w_1 + \frac{2P(2)}{\avg{k}}w_2(1-(1-w_1)^2) +(G_1(w_2)- \frac{P(1)}{\avg{k}} -\frac{2P(2)}{\avg{k}}w_2)(1-\tilde{G}_0(1-w_1))\right]\nonumber\\
	& & +(1-p)G_{1}(w_2)\left[1-G_{0}(1-{w}_1)\right] \\ 
	\hat{w}_2&=&p\left[\frac{P(1)}{\avg{k}}w_2 + \frac{2P(2)}{\avg{k}}(w_1+w_2^2(1-w1)) + 1-\frac{P(1)}{\avg{k}}-\frac{2P(2)}{\avg{k}}\right.\nonumber \\
	&&\left.-(G_1(1-w_1)- \frac{P(1)}{\avg{k}} -\frac{2P(2)}{\avg{k}}(1-w_1))(1-\tilde{G}_0(w_2))\right]\nonumber\\
	& & +(1-p)\left[1-G_{1}(1-w_1)+G_1(1-w_1)G_0\left(w_2\right)\right]
	\eea
\end{subequations}
where 
\bea
{G}_0(z)&=&\sum_{k}{P(k)}z^k\nonumber \\
{G}_1(z)&=&\sum_{k}\frac{k}{\avg{k}}P(k)z^k\nonumber  \\
\tilde{G}_0(z)&=&\sum_{k\ge 3}\frac{P(k)}{C}z^k.
\eea

Finally the energy $E$ is given by 
\bea
E&=&2\left\{G_0\left(\hat{w}_2\right)-\left[1-G_{0}(1-\hat{w}_1)\right]\right\}+2{\avg{k}}\left[\hat{w}_{1}(1-w_2)+w_{1}(1-\hat{w}_2)\right]-2(1-p)\left\{ [1-G_{0}(1-w_1)][1-G_{0}(w_2)]\right\}\nonumber \\
&-&2p\left\{ P(1)w_1(1-w_2)+P(2)(1-(1-w_1)^2)(1-w_2^2)+C(1-\tilde{G}_0(1-w_1))(1-\tilde{G}_0(w_2))\right\}
\eea

In the second case we consider the joint degree distribution $P^{in}(k^A,k^B)$ given by
\bea
P^{in}(k^A,k^B) =p\delta_{k^B,k^A}P(k^A)+(1-p)P(k^A)P(k^B),
\eea
where $P(k)$ is a given normalized degree distribution. 
The distributions of the fields over the links of this  ensemble of duplex networks are given by
\bea
{\cal P}_{\alpha}(h^{\alpha})&=&w_1^{\alpha}\delta(h^{\alpha}-1)+w_2^{\alpha}\delta(h^{\alpha}+1)+w_3^{\alpha}\delta(h^{\alpha}),\nonumber \\
\hat{{\cal P}}_{\alpha}(\hat{h}^{\alpha})&=&\hat{w}_1^{\alpha}\delta(\hat{h}^{\alpha}-1)+\hat{w}_2^{\alpha}\delta(\hat{h}^{\alpha}+1)+\hat{w}_3^{\alpha}\delta(\hat{h}^{\alpha}),
\eea
where $\alpha=A,B$  and where the probabilities $w_1^{\alpha},w_2^{\alpha},w_3^{\alpha}$ are normalized $w_1^{\alpha}+w_2^{\alpha}+w_3^{\alpha}=1$ as well as the probabilities $\hat{w}_1^{\alpha},\hat{w}_2^{\alpha},\hat{w}_3^{\alpha}$ that satisfy the equation $\hat{w}_1^{\alpha}+\hat{w}_2^{\alpha}+\hat{w}_3^{\alpha}=1$.
We get the equations
\begin{subequations}
	\bea
	\hat{w}_1&=&p\left[  G_1(w_2)-(1-w_1)G_1(w_2(1-w_1))  \right]+(1-p)G_{1}(w_2)\left[1-G_{0}(1-{w}_1)\right] \\ 
	\hat{w}_2&=&p\left[1-G_1(1-w_1)+w_2G_1(w_2(1-w_1))    \right]+(1-p)\left[1-G_{1}(1-w_1)+G_1(1-w_1)G_0\left(w_2\right)\right],
	\eea
\end{subequations}
where the generating functions $G_0(z)$ and $G_1(z)$ are defined as

\bea
{G}_0(z)=\sum_{k}{P(k)}z^k,\nonumber \\
{G}_1(z)=\sum_{k}\frac{k}{\avg{k}}P(k)z^k.
\eea

\begin{figure}[!t]
	\begin{center}
		{\includegraphics[width=0.99\columnwidth]{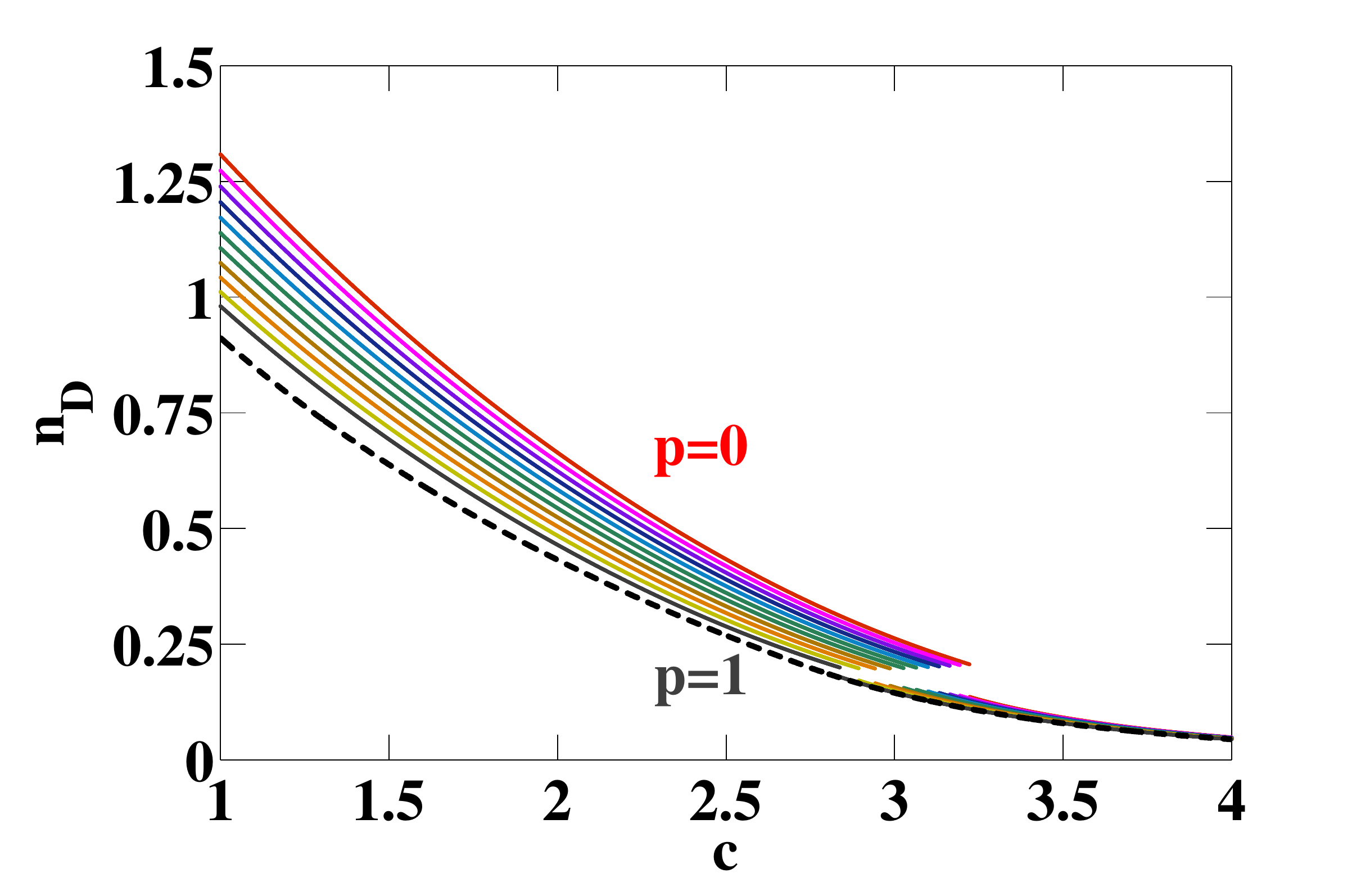}}
	\end{center}
	\caption{The density of the driver nodes $n_D$  in a duplex network formed by two Poisson networks with  $\Avg{k^{A,in}}=\Avg{k^{A,out}}=\Avg{k^{B,in}}=\Avg{k^{B,out}}=c$  and with correlated low in-degrees is plotted as function of $c$ for different values of  $p$. The result for the two separate layers is shown in black (dashed curve) while the situation for uncorrelated layers is shown in red. The value of $p$ increases going from the red curve ($p=0$) to the grey curve ($p=1$).} 
	\label{lowdegreefullcolor}
\end{figure}

The energy $E$ in this ensemble is given by 
\bea
E&=&2\left\{G_0\left(\hat{w}_2\right)-\left[1-G_{0}(1-\hat{w}_1)\right]\right\}+2{\avg{k}}\left[\hat{w}_{1}(1-w_2)+w_{1}(1-\hat{w}_2)\right]\nonumber \\
&-&2(1-p)\left\{ [1-G_{0}(1-w_1)][1-G_{0}(w_2)]\right\}-2p\left\{ 1-G_0(1-w_1)-G_0(w_2)+G_0(w_2(1-w_1))\right\}
\eea
The degree correlation of low in-degree nodes can modify the number of driver nodes $n_D$ found in duplex networks (see Fig. \ref{lowdegreefullcolor} for the case of a duplex network formed by Poisson layers with  $\Avg{k^{A,in}}=\Avg{k^{A,out}}=\Avg{k^{B,in}}=\Avg{k^{B,out}}=c$).
Once the low in-degree nodes are correlated, correlating also the other in-degrees of the network does not change substantially the number of driver nodes $n_D$ as discussed in the main body of the paper.

\end{document}